\documentclass[fleqn,usenatbib]{mnras}

\usepackage{newtxtext,newtxmath}

\usepackage[T1]{fontenc}
\usepackage{ae,aecompl}

\usepackage{graphicx}	
\usepackage{amsmath}	
\usepackage{amssymb}	

\usepackage[T1]{fontenc}
\usepackage{ae,aecompl}

\usepackage{bm}
\usepackage{comment}
\usepackage{color}
\usepackage[caption=false]{subfig}
\graphicspath{ {./Figures/} }

\newcommand{\vpeak}{v_{\textrm{peak}}}
\newcommand{\mpch}{$h^{-1}$Mpc}
\newcommand{\msolarh}{$h^{-1}$M$_{\odot}$}
\newcommand{\galform}{{\scshape galform}}
\newcommand{\lgalaxies}{{\scshape l-galaxies}}
\newcommand{\subfind}{{\scshape subfind }}
\newcommand{\Mh}{M_{\rm h}}
\definecolor{ForestGreen}{rgb}{0.3,0.7,0.3}

\title[HODs revisited]{Revisiting HOD model assumptions: the impact of AGN feedback and assembly bias}
\author[N. McCullagh et al.]{
Nuala McCullagh,$^{1}$\thanks{E-mail: nuala.mccullagh@durham.ac.uk. The Python code used to generate the figures in this paper can be found at \url{https://github.com/nualamccullagh/hods-revisited}}
Peder Norberg,$^{1,2}$
Shaun Cole,$^{1}$
Violeta Gonzalez-Perez,$^{3}$
\newauthor{
Carlton Baugh,$^{1}$
and John Helly$^{1}$
}
\\
$^{1}$Institute for Computational Cosmology, Department of Physics, Durham University, South Road, Durham DH1 3LE, UK\\
$^{2}$Centre for Extragalactic Astronomy, Department of Physics, Durham University, South Road, Durham DH1 3LE, UK\\
$^{3}$Institute of Cosmology and Gravitation, University of Portsmouth, Dennis Sciama Building, Portsmouth PO1 3FX, UK
}

\date{Accepted XXX. Received YYY; in original form ZZZ}

\pubyear{2017}

\begin{document}
\label{firstpage}
\pagerange{\pageref{firstpage}--\pageref{lastpage}}
\maketitle

\begin{abstract}
The standard form of Halo Occupation Distribution (HOD) models were originally developed based on results from semi-analytic and hydrodynamical galaxy formation models. Those models have since progressed, in particular to include AGN feedback to match the galaxy luminosity function in a universe with the observed baryon fraction. AGN feedback affects the relationship between galaxy stellar mass and luminosity, in particular making the relationship non-monotonic. For matched number density samples, galaxies in luminosity-threshold samples occupy a different range of halo masses from those in stellar-mass-threshold samples. We find that the shapes of the HODs of luminosity-threshold samples are slightly more complicated in semi-analytic galaxy formation models that include AGN feedback than are assumed by standard HOD models, even when the large-scale clustering matches. We also find that subhalo abundance matching (SHAM) does not preserve these non-standard shapes. We show that catalogues created using SHAM and the semi-analytic model \galform\ that have the same large-scale 2-point clustering by construction have different void probability functions (VPFs) in both real and redshift space. We find that these differences arise from the different HOD shapes, as opposed to assembly bias, which indicates that the VPF could be used to test the suitability of an HOD model with real data.
\end{abstract}




\section{Introduction}
\label{sec:intro}

The statistical model known as the Halo Occupation Distribution 
\citep[HOD][]{Seljak2000,Peacock2000,berlind2002}, which built on the
study of galaxy clustering \citep{kauffmann1997,benson2000}
in semi-analytic models of galaxy formation,
has become an important tool in connecting the observed galaxy
distribution to the underlying distribution of dark matter. 
It has two distinct applications. In the forward direction 
it is used to produce mock galaxy catalogues from dark matter only
N-body simulations \citep[e.g.][]{Smith2017}
or other approximate methods of generating dark
matter halo catalogues. The resulting mock galaxy catalogues are now
recognized as a vital part of all galaxy redshift surveys. They are used for
survey design and forecasts, in developing and testing analysis tools,
and to provide Monte Carlo estimates of covariance matrices in order to
obtain robust parameter constraints \citep{Manera2013}. In the reverse direction the HOD is
used to model the observed galaxy clustering and infer the statistics
of how galaxies populate dark matter haloes \citep[e.g.][]{zehavi2011,RodriguezTorres2016}, which in turn constrains
galaxy formation models.

The HOD model makes the explicit assumption that the probability that
a dark matter halo hosts a galaxy with particular properties depends
only on the mass of the halo. Thus for galaxies specified by a
threshold in stellar mass or luminosity the HOD is completely
specified by the probability distribution $P(N|M_{\rm h})$, the probability
that a halo of mass $M_{\rm h}$ hosts $N$ such galaxies. In turn this
distribution is normally specified by a simple analytic functional
form \citep[e.g.][]{Zheng2005} for how the mean of the distribution,
$\langle N(M_{\rm h}) \rangle$, depends on halo mass (see \S2.3) and a simple ansatz for the
dispersion about this mean.\footnote{$\langle N(M_{\rm h}) \rangle$ is 
decomposed into central, $\langle N_{\rm cen}(M_{\rm h}) \rangle$,
and satellite, $\langle N_{\rm sat}(M_{\rm h}) \rangle$, contributions.
As there can be only one or zero central galaxies the fraction of haloes with one central galaxy simply equals
$\langle N_{\rm cen}(M_{\rm h}) \rangle$.
Normally the satellite occupation distribution is assumed to be Poissonian.}

To be confident in the inferences that are made from HOD
modelling it is important to scrutinize the assumptions that underpin
it. The functional forms used in standard HOD models were originally motivated
by the form found in semi-analytic models and hydrodynamical galaxy
formation simulations \citep{berlind2003}. However, since then it has been recognised that AGN play a much larger role in modulating the cooling of gas in massive haloes \citep{bower2006,croton2006} than was assumed in that
generation of models and simulations. Hence it is interesting to
compare the clustering and HODs of current semi-analytic models with 
that assumed in most HOD modelling. We investigate deviations of the
semi-analytic HOD from the assumed form and whether these deviations
have a measurable effect on galaxy clustering. Even if the HOD is
described well by a standard analytic form, the clustering in the 
semi-analytic model could differ from that of the HOD prediction due
to assembly bias, which is the degree to
which the formation and properties of a halo depend on its environment. 
Specifically semi-analytic models make full use of
the merger history of each dark matter halo and these are known to
depend on environment as well as mass \citep{Gao2007}.

An alternative way of placing galaxies in dark matter cosmological N-body simulations is the method of sub-halo abundance matching
\citep[SHAM][]{conroy2006,vale2006,hearin2016}. This method, described in \S \ref{sec:SHAM}, does not assume that the number of galaxies hosted by a
 particular halo depends only on the halo mass. Halos of the same mass 
can have differing numbers of subhaloes and the way in which this
depends on formation history and environment is captured
directly in the N-body simulation. Hence assembly bias, is included, at least to some extent, in the SHAM process, but neglected in HOD models. To investigate the differences in clustering which are due to assembly bias and those that are due to 
the effects of the non-standard HOD, we compare the clustering of our
semi-analytic model with two other catalogues. Firstly one produced by applying the SHAM process to the semi-analytic catalogue, resulting in one that still contains assembly bias, but has a more standard HOD shape. Secondly we compare to a catalogue where galaxies are
shuffled between haloes of the same mass which preserves the
non-standard HOD, but erases assembly bias. In this way we can 
disentangle the effects of the non-standard HOD from assembly bias
and identify their distinctive characteristics.

The paper is organized as follows: in \S\ref{sec:sims} we describe the N-body simulation used throughout this paper, as well as the methods used
to populate dark matter haloes with galaxies, namely \galform, SHAM, and HOD. In \S\ref{sec:clustering}, we present
the clustering and measured HODs of \galform\ luminosity-threshold samples with the same number densities as SDSS samples given in \citet{zehavi2011}. We compare the \galform\ samples to both SDSS and to SHAM samples made to match the large-scale clustering of \galform. In \S\ref{sec:discussion}, we investigate whether we can distinguish samples with the same large-scale clustering and different HOD shapes, and find that the void probability function (VPF) can be used for this purpose. We conclude in \S\ref{sec:conclusion}.

\section{Simulations and Methods for Creating Galaxy Catalogues}
\label{sec:sims}

We use the subhalo merger trees from the P-Millennium simulation \citep{baugh_pmill}, which is a dark-matter only $N$-body simulation with Planck cosmology: $H_0=67.77$ km~s$^{-1}$~Mpc$^{-1}$, $\Omega_{\Lambda}=0.693$, $\Omega_{\rm M}=0.307$, $\sigma_8=0.8288$ \citep{planck_cosmology}. The box size is $L_{\rm b}=800$ Mpc or $542.16$ \mpch, and particle mass $M_{\rm p}=1.061 \times 10^{8}$ \msolarh\ ($5040^3$ dark matter particles). Between the starting redshift, $z_{\textrm{init}}=127$, and the present day, $z=0$, particle data was written out at 272 snapshots, of which 16 have been saved and the others were discarded after (sub)halo catalogues were created from them.

Subhalo merger trees are constructed using the P-Millennium snapshots. Haloes are identified in each snapshot using the friends-of-friends algorithm, and then subhaloes are identified using \subfind \citep{springel2001}. Subhaloes are tracked between output times, and a new set of haloes (called Dhaloes) are created that have consistent membership over time \citep[see Appendix A of][for more details]{jiang2014}. 

We use two methods to populate the P-Millennium subhalo catalogue with galaxies: \galform\ and SHAM. We compare the clustering of the resulting samples with that of luminosity threshold samples from SDSS by \citet{zehavi2011}. We also compare the HODs of these galaxy catalogues, which we measure directly from the simulations, to those inferred by \citet{zehavi2011} through HOD modelling of the observed SDSS galaxy clustering.

\subsection{GALFORM}
\label{sec:galform}

Semi-analytic galaxy formation models such as \galform\ model the formation and evolution of galaxies using simple, physically motivated equations to predict baryonic physics within dark matter halo merger trees \citep[e.g.][]{cole2000, baugh2005, bower2006, galform2014, lacey2016}. We use simulated galaxy samples from \galform\ 
run on the P-Millennium simulation \citep{baugh_pmill}. The output is a galaxy catalogue with realistic clustering. We use the snapshot at $z=0.1$ to match the effective redshift of the samples in \citet{zehavi2011}.

The \galform\ model used here is an evolution of the model described in \citet{galform2014} and \citet{galeagle2016}. Like \citet{gonzalez2017}, it includes a new treatment of merging satellites \citep{Simha2016}. In essence this model is very similar to that of \citet{gonzalez2017}, except for the cosmology considered, with the latter using WMAP7 cosmology with the associated MR7 N-body simulation \citep{guo13}. Hence the model parameters have been recalibrated to account for the new merging scheme, the higher resolution N-body simulation and the new cosmology compared to \citet{galform2014} and \citet{gonzalez2017}: in this model, both supernova feedback and AGN feedback are slightly less efficient. We refer the reader to both \citet{gonzalez2017} and \citep{baugh_pmill} for more specific \galform\ details regarding this model and the tuning process of the model parameters.

As for the satellite merging scheme, previous \galform\ models used analytic dynamical friction models \citep{colelacey93,Jiang2008} to follow the orbits of satellite galaxies. In the current model, the subhaloes hosting galaxies are followed until they are stripped and destroyed, or until they merge with another subhalo. When a galaxy loses its host subhalo (not due to a subhalo merger), it follows the previously most bound particle and its fate is determined with an analytic formula tuned to produce good agreement between subhalo abundances in Millennium I and the 125 times higher-resolution Millennium II simulations \citep[see][for more details]{Simha2016}. This work is the first \galform\ study presenting clustering properties of \galform\ galaxies at $z\sim 0.1$ since the adoption of the new satellite merging scheme, while \citet{gonzalez2017} considers that of emission line galaxies at $z\sim 1$.  

\begin{figure}
\centering
\subfloat{
\includegraphics[width=0.48\textwidth]{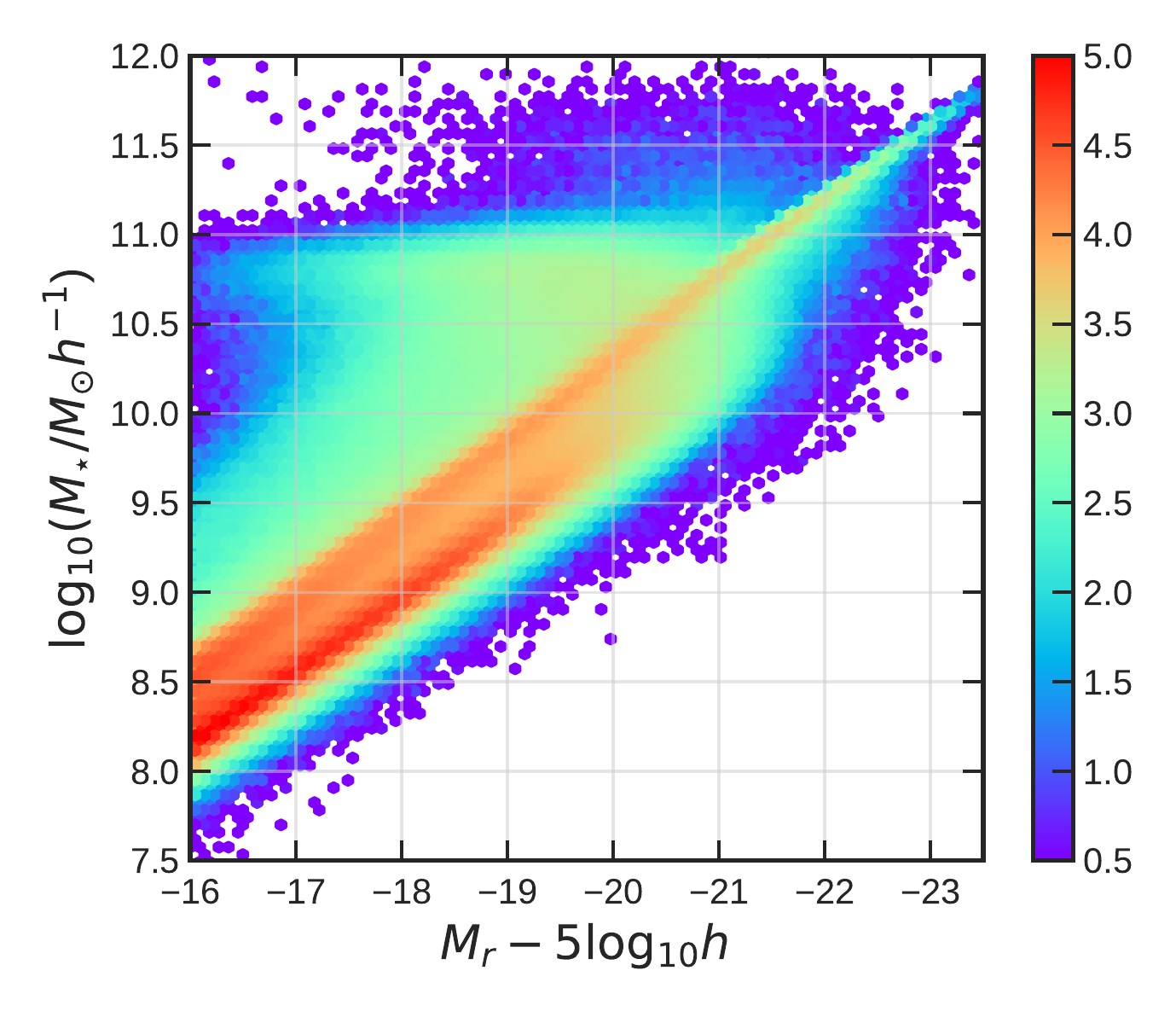}
}
\\
\subfloat {
\includegraphics[width=0.48\textwidth]{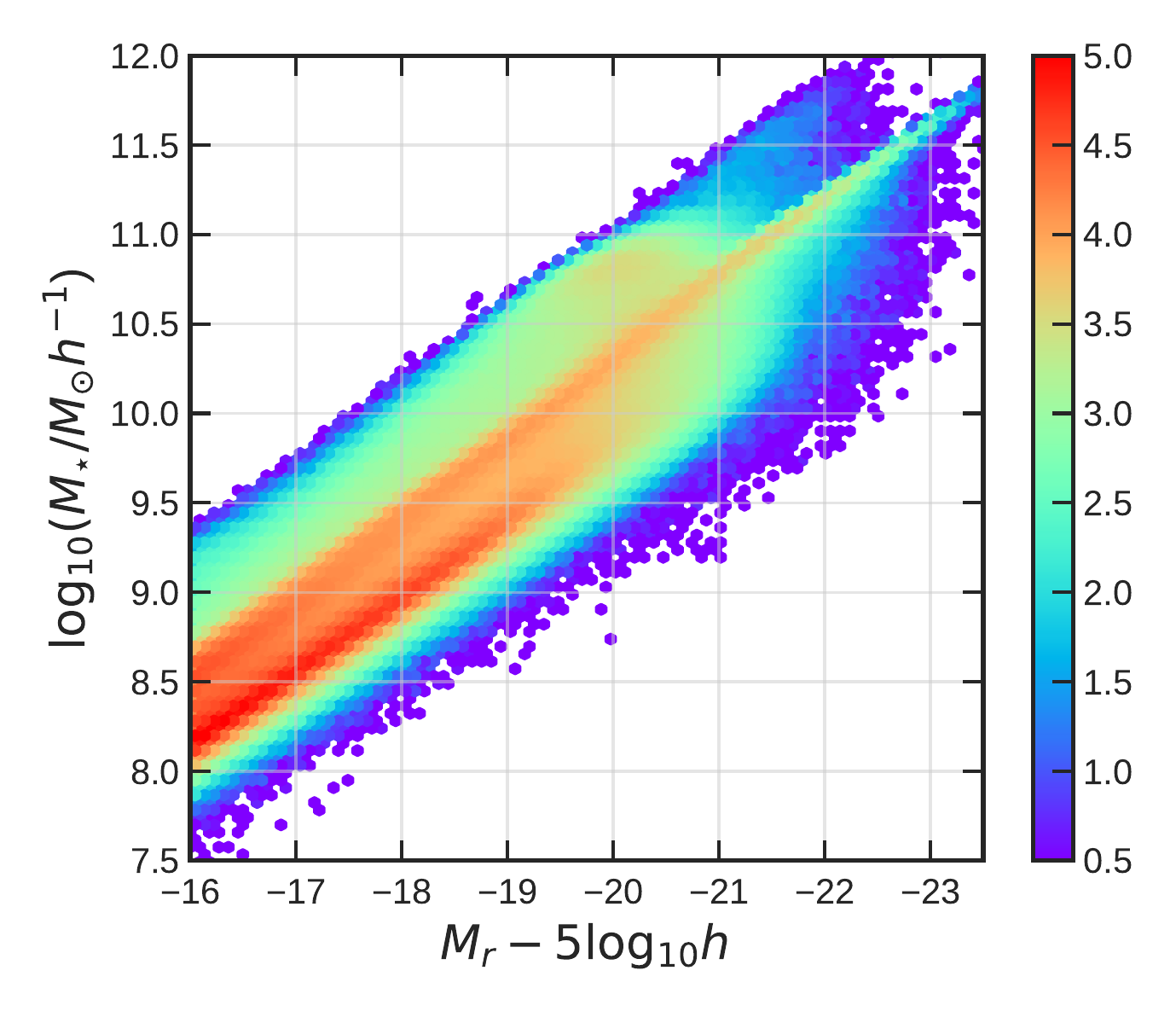}
}
\caption{Stellar mass versus r-band magnitude of \galform\ galaxies. The top panel shows the magnitude with the full dust extinction computed by \galform. The bottom panel shows the magnitude with the tapered dust extinction. The colour shows the $\log_{10}$ of the number of galaxies in each 2D bin.}
\label{fig:gal_mstar_magr}
\end{figure}

\begin{figure}
\begin{center}
\includegraphics[width=0.48\textwidth]{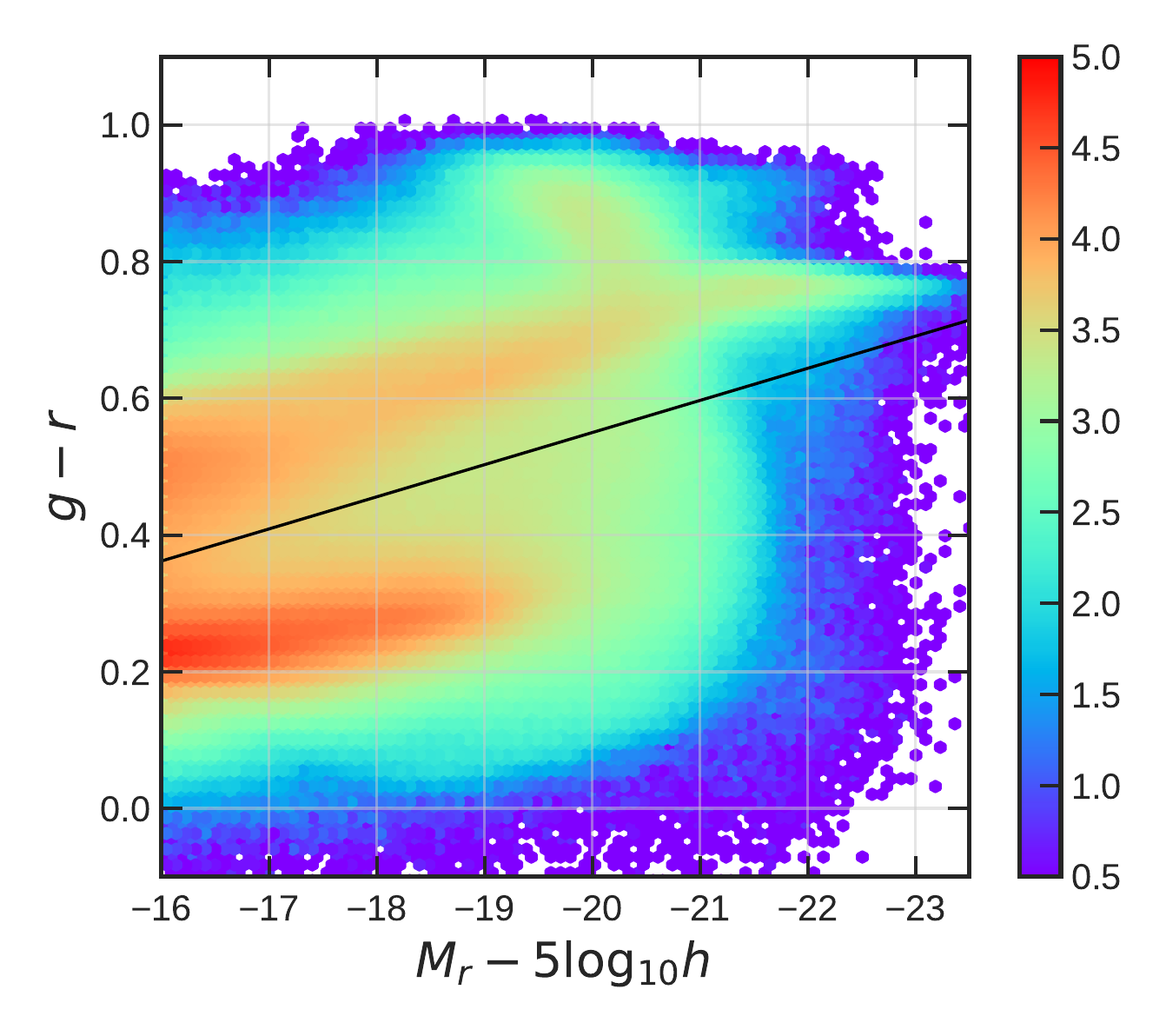}
\caption{Colour-magnitude diagram of \galform\ galaxies with tapered dust extinction included. The colour scale is the same as in Fig.~\ref{fig:gal_mstar_magr}. The black line shows a colour cut separating red and blue galaxies. }
\label{fig:gal_col}
\end{center}
\end{figure}

One aspect of \galform\ that is relevant to our study is the treatment of dust attenuation. Because we are defining our samples based on r-band luminosity, we have the choice in \galform\ to use the total magnitude, or the magnitude with dust extinction included\footnote{Throughout we refer to absolute magnitude as ``magnitude" only.}. We note that \galform\ is tuned to match the luminosity function of observed galaxies including dust extinction. We find that including dust is important in achieving accurate luminosity-dependent clustering, but that a small fraction of the extinctions computed by \galform\ are unphysically large (e.g. about 7\% of the galaxies have greater than 2.0 magnitudes of extinction in the faintest sample we consider). The way \galform\ currently computes dust attenuation is related to the sizes of the galaxies, which do not accurately match observations and contain a small population of unrealistically compact galaxies \citep[see][]{lacey2016}. Due to their sizes, these galaxies end up with unrealistically large optical depths, and therefore large extinctions \citep{gp2013,lacey2016, merson2016}. We therefore apply a tapering to the r-band extinction which preserves the ordering (the galaxies with the largest extinction remain so) but imposes a maximum value of the extinction. The tapering has a negligible impact on the overall dust extincted luminosity function. The tapering procedure is described in detail in Appendix \ref{sec:appendix}. The top panel of Fig.~\ref{fig:gal_mstar_magr} shows the stellar mass versus r-band luminosity in \galform\ including full dust extinction, and the bottom panel shows this relation with the tapered dust extinction, with maximum extinction set to 2.0 magnitudes. This figure shows that the population of galaxies that are given unphysically large extinctions in \galform\ have stellar masses that are typically > $10^{10.5}$ \msolarh, and that they are spread over a range of luminosities. 

We find that the tapering does not significantly affect the clustering or HOD of any of our \galform\ samples as compared to samples made with un-tapered magnitudes, because the highly extincted galaxies are always a negligible contribution to the population in any luminosity threshold sample. In the remainder of this paper, the \galform\ r-band magnitudes used include dust extinction tapered to a maximum of 2.0 magnitudes, and g-band magnitudes are tapered consistently as described in Appendix \ref{sec:appendix}. While the choice of 2.0 magnitudes as a maximum value is somewhat arbitrary, we note that tapering to this value has little effect on the majority of the galaxies: in the faintest sample we consider, roughly 10\% of the extinctions are changed by more than 0.4 magnitudes by the tapering procedure.

Fig.~\ref{fig:gal_col} shows the (tapered) colour-magnitude diagram of \galform\ galaxies. The clouds of red and blue galaxies are clear in this figure, and we show an empirical colour cut (black line) corresponding to: $(g-r) = -(M_r-20)/21.28 + 0.55$. This is the cut used to separate red and blue galaxies later in the paper. 

\begin{figure}
\centering
\subfloat{
\includegraphics[width=0.48\textwidth]{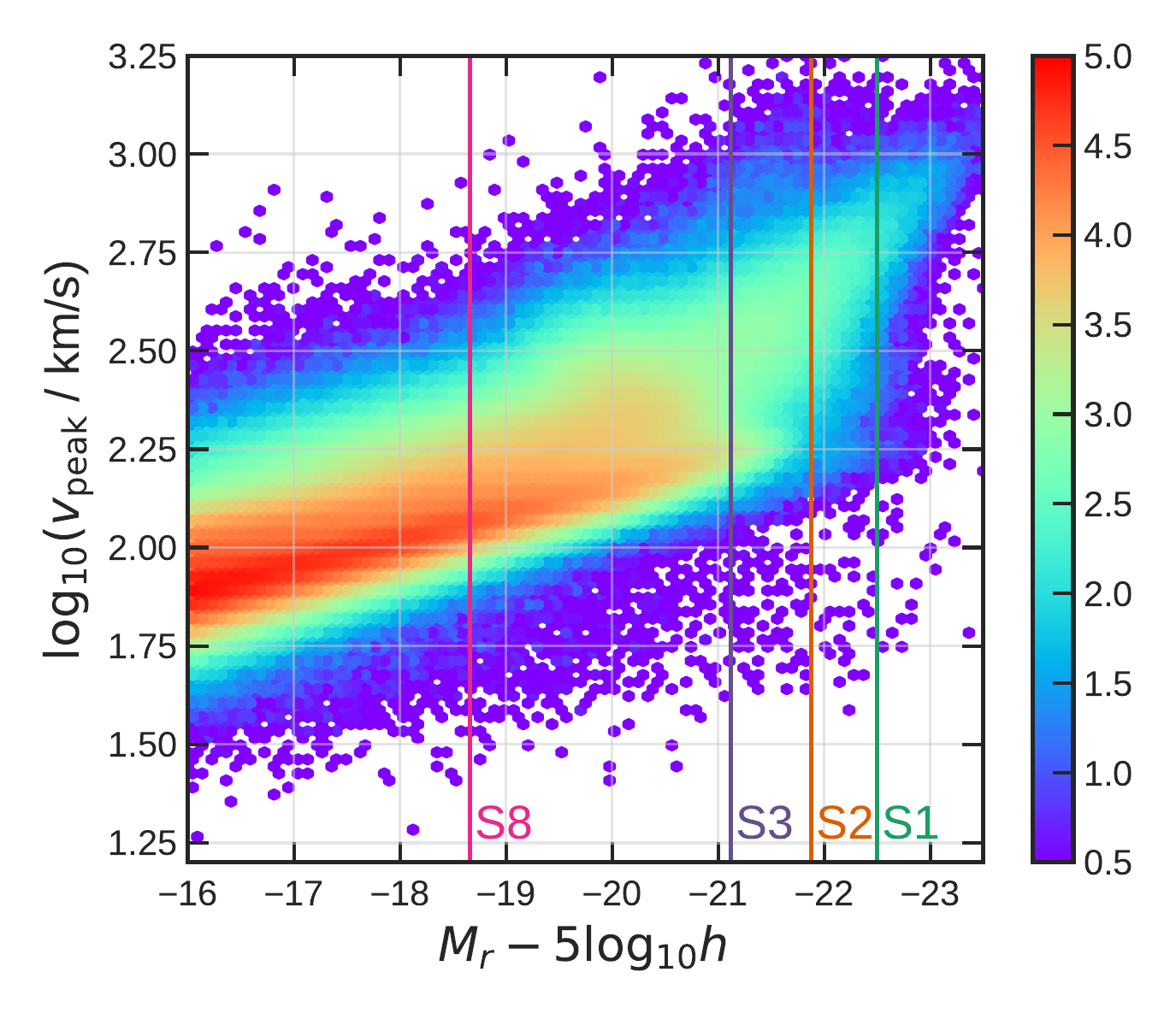}
}
\\
\subfloat {
\includegraphics[width=0.48\textwidth]{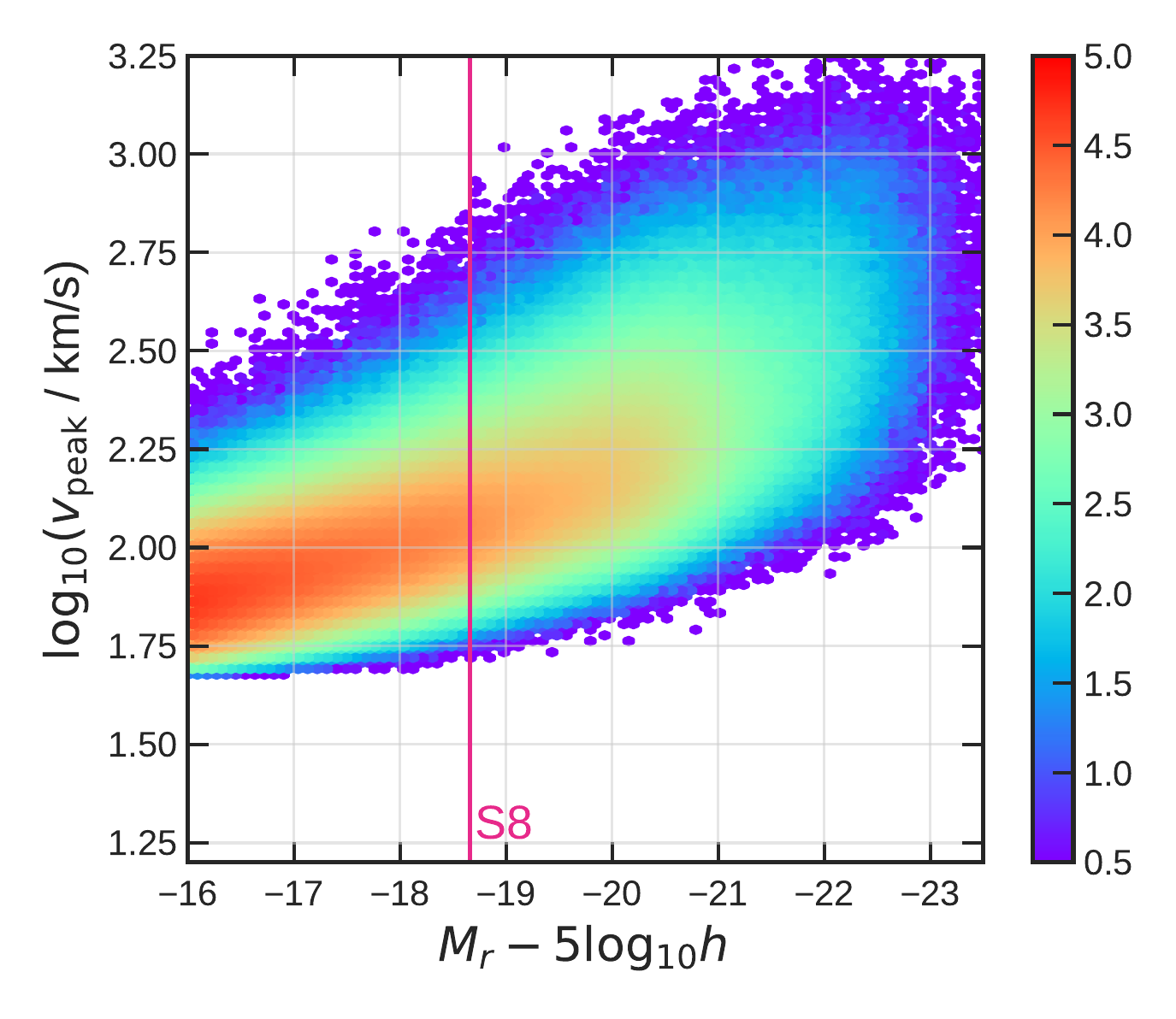}
}
\caption{Top panel: subhalo peak velocity ($\vpeak$) versus r-band magnitude in \galform. Bottom panel: $\vpeak$ versus r-band magnitude for a SHAM catalogue with $\Delta M_r=1.2$. In the top panel, the vertical lines correspond to the luminosity cuts for the samples used throughout the paper. A scatter of $\Delta M_r=1.2$ is what is needed to create a SHAM catalogue that matches the clustering of Sample 8, whose luminosity threshold is indicated by the vertical line in the lower panel. For this SHAM catalogue, only objects to the right of this line are used. The colour scale is the same as in Fig.~\ref{fig:gal_mstar_magr}.}
\label{fig:vpeak_mr}
\end{figure}

\subsection{Subhalo Abundance Matching}
\label{sec:SHAM}

The other method we use to populate subhaloes with galaxies is SubHalo Abundance Matching (SHAM). In SHAM, luminosities or stellar masses from an observed sample of galaxies are mapped onto a subhalo catalogue by requiring the number of galaxies above a certain luminosity or mass threshold to be the same as the number of subhaloes above a threshold in some property of the subhalo. In its simplest form, SHAM requires only one free parameter: the choice of what subhalo property is mapped onto the galaxy luminosity or stellar mass. Following common practice, we use the quantity $v_{\textrm{peak}}$, which is the peak value that the maximum circular velocity attains along a subhalo's merger history \citep[e.g.][]{Reddick2013}. We focus here on the luminosity as opposed to stellar mass because observationally the samples are best defined in terms of luminosity, as they are flux limited by nature. SHAM then takes the following form:
\begin{equation}
n_g(>L) = n_h(>\vpeak).
\end{equation}
Because we match on the quantity $\vpeak$, which contains some information about the subhalo merger history, we expect that our SHAM catalogue will have some level of assembly bias \citep{croton2007, zentner2014, hearin2015}.

In practice, it is necessary to introduce scatter in this relation for the large-scale clustering of the SHAM catalogue to match that of the galaxy sample, which introduces at least one additional free parameter per galaxy sample into SHAM. In this work, we introduce a fixed scatter in magnitude ($\Delta M_r$) for a given sample in the following way:
\begin{itemize}
\item From the array of galaxy magnitudes, $M_r$, create an array, $M_r'$, which are each drawn from a Gaussian distribution with mean equal to the initial $M_r$ and fixed $\sigma=\Delta M_r$
\item Find the ordering that will rank order the new array of scattered magnitudes ($M_r'$), and sort the array of (original) galaxy magnitudes ($M_r$) by that ordering
\item Rank order $\vpeak$ of the subhaloes
\item Assign the original values of galaxy magnitudes ($M_r$) to the subhaloes such that the subhalo with the highest $\vpeak$ gets assigned the galaxy magnitude with the brightest $M_r'$
\end{itemize}

In this work we take the r-band magnitudes from \galform\ as our `observed sample of galaxies' and use abundance matching to assign the magnitudes to subhaloes. We add scatter to match the observed large-scale clustering in \galform\ for each sample, which implies that in our implementation, each sample requires a different level of scatter.

Note that we perform abundance matching on the P-Millennium subhalo catalogue at a single snapshot, which means that our SHAM catalogues do not include orphan galaxies \citep{guowhite2014}, whereas \galform\ contains orphan galaxies by tracking galaxies that lose their subhalo. The fraction of orphans in each of our samples can be found in Table \ref{tab:samples}.

\begin{figure*}
\centering
\subfloat{
\includegraphics[width=0.48\textwidth]{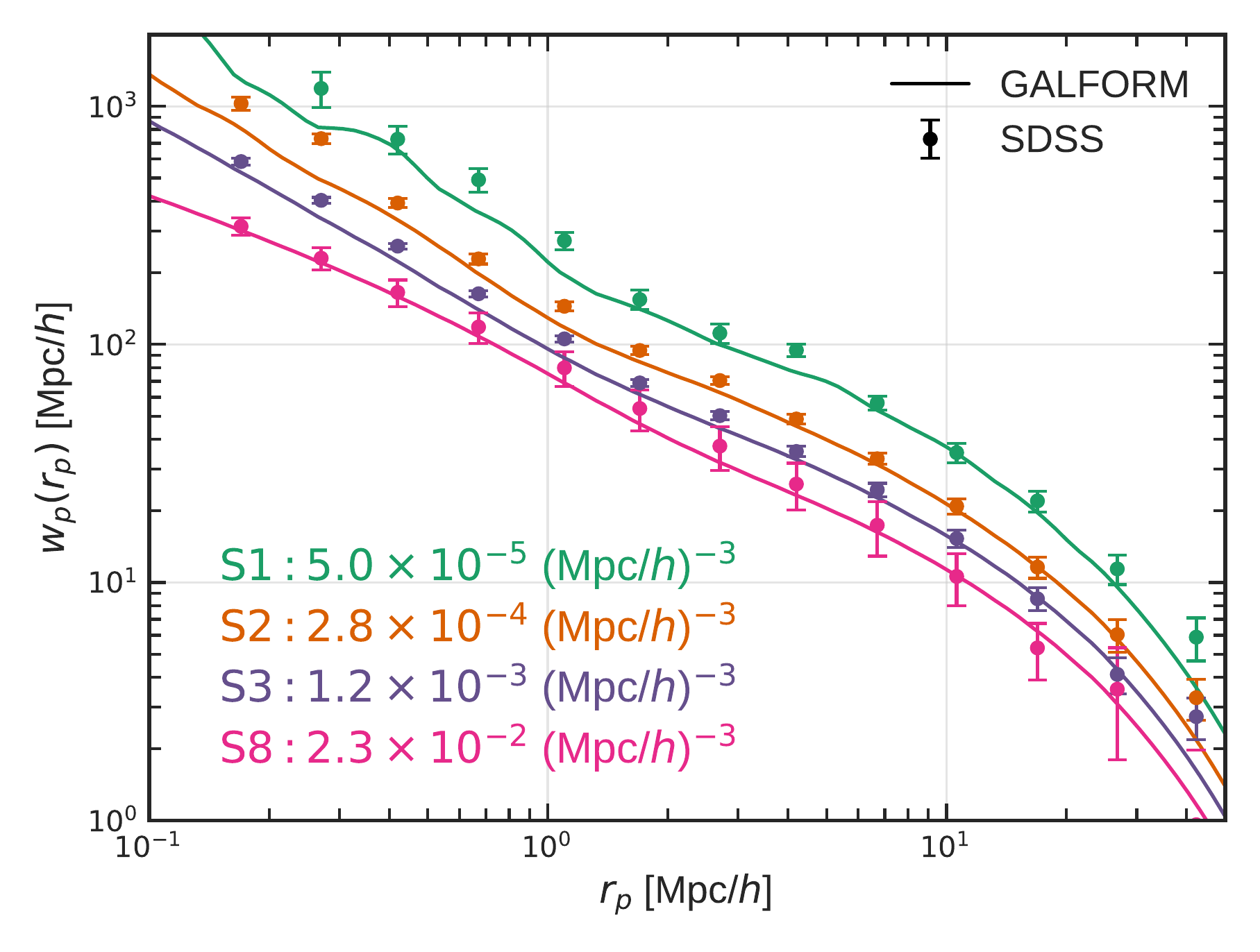}
}
\subfloat {
\includegraphics[width=0.48\textwidth]{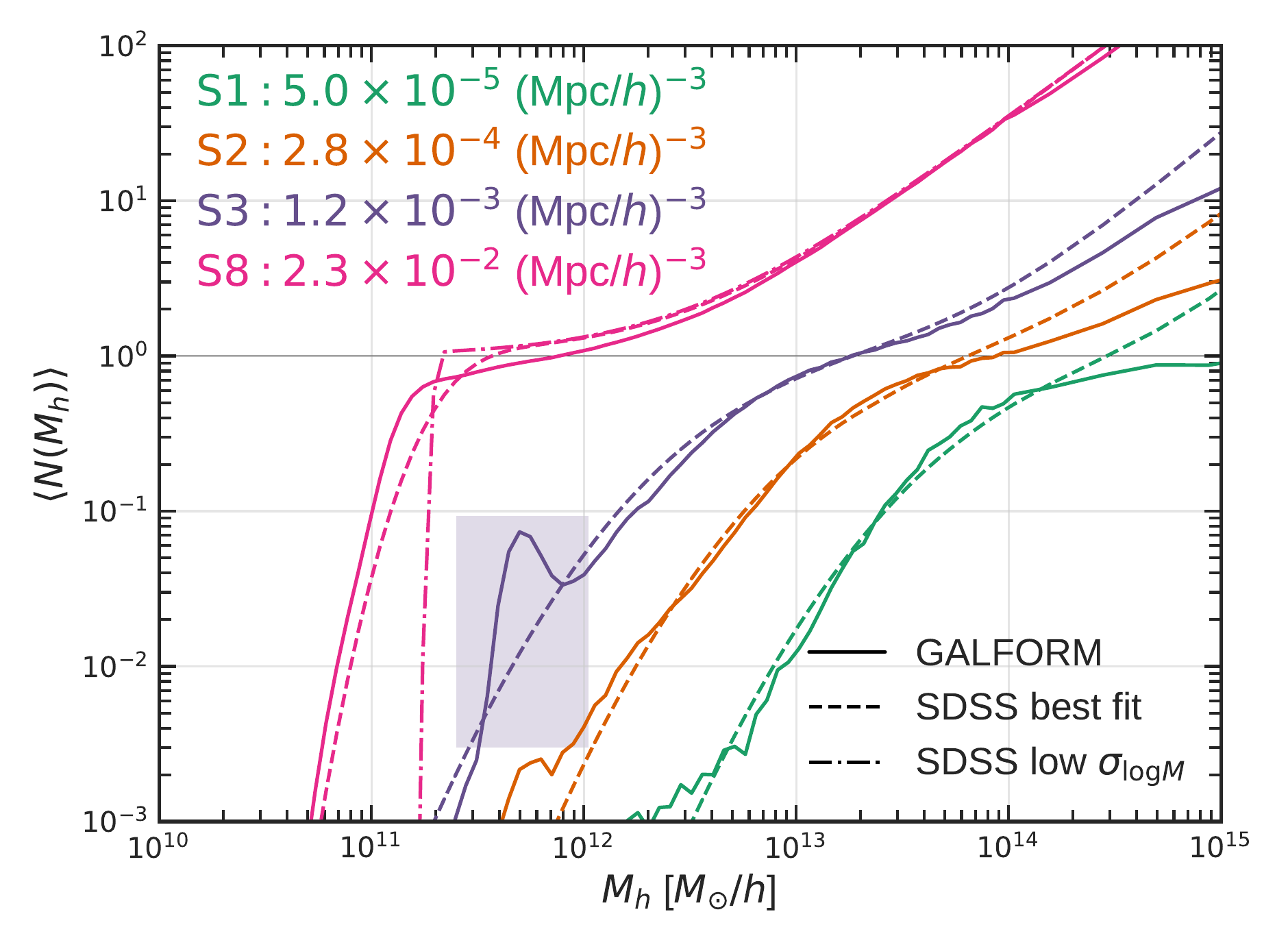}
}
\caption{Left panel: projected correlation function of \galform\ (solid lines) versus the corresponding sample in SDSS (points with error bars). Right panel: the corresponding HODs of the samples. These are measured directly from the simulation for the \galform\ samples, and inferred through HOD fitting to the clustering for SDSS. All halo masses are mapped onto the WMAP7 \citet{jenkins2001} mass function at $z=0.1$ to ensure the HODs can be compared on a fair basis. The shaded box highlights the feature in S3 that we attribute to AGN feedback.}
\label{fig:galformSDSS}
\end{figure*}

By matching the \galform\ galaxies to their true subhaloes in the subhalo merger trees, We can compare the true underlying properties in the \galform\ catalogue to those inferred by SHAM. For example, the top panel in Fig.~\ref{fig:vpeak_mr} shows the true relationship between $v_{\textrm{peak}}$ and r-band magnitude in \galform. The vertical lines correspond to the luminosity cuts for the 4 samples we highlight later in the paper (S1, S2, S3, and S8 in Table~\ref{tab:samples}). The bottom panel of Fig.~\ref{fig:vpeak_mr} shows the resulting $v_{\textrm{peak}}$ -- r-band magnitude relation in a SHAM catalogue with scatter. For a SHAM catalogue with no scatter, this relation would be a line. The SHAM catalogue shown in Fig.~\ref{fig:vpeak_mr} uses a scatter of $\Delta M_r=1.2$, which is the amount needed to match the clustering in Sample 8 in Table~\ref{tab:samples}, and the luminosity cut for this sample is indicated by the vertical line in the lower panel. Only objects to the right of this line are used in this SHAM catalogue. As Fig.~\ref{fig:vpeak_mr} shows, the $v_{\textrm{peak}}$ -- r-band magnitude relation in the SHAM catalogue is much simpler and contains less structure than that of \galform, which is to be expected from the simplistic procedure of abundance matching.

\subsection{Halo Occupation Distribution}
\label{sec:HOD}

The final method we consider for connecting galaxies to their haloes is the Halo Occupation Distribution (HOD) model. While a HOD can itself be used to populate a halo catalogue with galaxies given some assumptions about the density profile of the halos, here we use the HOD purely as a descriptor of the samples in our galaxy catalogues.

For a given sample, a full HOD model specifies the probability that $N$ galaxies live in a halo of a given mass: $P(N | M_h)$. This conditional probability is often assumed to be Poisson with mean $\langle N(M_h) \rangle$, which can be modelled in various ways. Here we consider the 5-parameter model described in \citet{zehavi2011} \citep[see also][]{Zheng2005, Zheng2007, Zheng2009} for the mean occupation function $\langle N(M_h) \rangle$:
\begin{align}
\left \langle N(M_h) \right \rangle &=\frac{1}{2}\left [ 1+\textrm{erf}\left ( \frac{\log M_h - \log M_{\textrm{min}}}{\sigma_{\log M}} \right )\right ]\notag\\
& \qquad \times \left [ 1+ \left ( \frac{M_h- M_0}{M_1'}\right )^{\alpha}\right ].\label{eq:hodmodel}
\end{align}
The first line in this expression is a step function which describes the occupation of central galaxies. It has mass scale $M_{\textrm{min}}$ and is smoothed with $\sigma_{\log M}$ to account for scatter between galaxy luminosity (or stellar mass) and halo mass. The second line in Eq.~\ref{eq:hodmodel} is a power law with index $\alpha$ in halo mass, which is associated with the occupation of satellite galaxies. The satellite term also has a cutoff mass scale $M_0$ and normalization $M_1'$. Note that this formulation of the HOD does not include any assembly bias, in that the number of galaxies in a halo is only a function of the halo mass, and nothing else. However, it is possible to extend standard HOD models to include dependences on other parameters \citep[e.g.][]{hearin2015,hearin2016}.

In simulations, there are a variety of ways to define the halo mass, $M_{\rm h}$. Here we use the DHalo mass from the \galform\ catalogue. This gives a halo mass function that agrees reasonably well with that predicted from \citet{jenkins2001} for the P-Millennium cosmology for $M_{{\rm m}, 200}$, which is the mass within $r_{{\rm m},200}$, the radius at which the mean enclosed density of the halo is 200 times the mean density of the universe. To compare our P-Millennium measurements to the SDSS analysis, we map our DHalo mass function onto a Jenkins mass function of $M_{{\rm m}, 200}$ in a WMAP7 cosmology at $z=0.1$. To map one mass function to another, we compute the cumulative number density for a given mass in one cosmology and associate it to the mass at the same number density in the other cosmology.

In the halo model, the mean occupation function can be connected to the bias and number density of a given galaxy sample \citep{smt2001} \citep[for a review, see][]{CooraySheth2002}.
The number density of a galaxy sample is defined as:
\begin{equation}
\bar n_{\rm g}=\int_0^{\infty}\frac{d \bar{n}_h}{d \Mh} \langle N(\Mh) \rangle \, d \Mh, \label{eq:numberdensint}
\end{equation}
and the bias of a sample is:
\begin{equation}
b_{\rm g} =
\frac{1}{{\bar n}_{\rm g}}\int_0^{\infty}
\frac{d{\bar n}_{\rm h}}{d \Mh} 
\langle N(\Mh)  \rangle 
b(\Mh)\, d \Mh ,
\label{eq:biasint}
\end{equation}
where $b(\Mh)$ is the halo bias as a function of halo mass, and $\frac{d \bar{n}_h}{d \Mh}$ is the halo mass function. We use the halo bias relation predicted from ellipsoidal collapse \citep[Eq.~8 in][]{smt2001}, and we use the \citet{jenkins2001} mass function, as this is what is used in \citet{zehavi2011}. Note that we do not use Eq.~\ref{eq:biasint} to determine the bias of our samples, as we measure the clustering directly, but we use both Eqs.~\ref{eq:numberdensint} and~\ref{eq:biasint} to examine the relationship between the HODs of our samples and their number density and their large-scale bias.

\section{GALFORM Clustering Results}
\label{sec:clustering}

In this section we discuss the clustering and HOD of our \galform\ catalogue compared to both SDSS \citep{zehavi2011} and SHAM. We limit our discussion to the results from \galform\ specifically, and we will generalize our findings in \S\ref{sec:discussion}. 

We create galaxy samples by rank ordering the (tapered) r-band magnitude and matching number densities from the samples in \citet{zehavi2011}. Because the luminosity function in \galform\ is slightly different from that in SDSS, and because we are using the rest-frame r-band magnitude from \galform\ at $z=0.1$, the luminosity thresholds of the samples differ slightly from those in SDSS. These are given in Table~\ref{tab:samples}. We will refer to the samples by their number densities as opposed to luminosity thresholds for clarity.

\subsection{Comparison: GALFORM vs SDSS}
\label{sec:galformSDSS}

We measure the real-space 2-point correlation functions of our samples using the {\tt CUTE} pair-counting code\footnote{The {\tt CUTE} code is publicly available at \url{http://members.ift.uam-csic.es/dmonge/CUTE.html}} \citep{cute}. In order to compare the clustering in \galform\ with that of SDSS from \citet{zehavi2011}, we compute the projected correlation function $w_p(r_p)$, which is related to the real-space correlation function $\xi(r)$, by:
\begin{equation}
w_{\rm p}(r_{\rm p}) = 2\int_{r_{\rm p}}^{\infty} \xi(r)(r^2-r_{\rm p}^2)^{-1/2}r\ dr.
\end{equation}
In practice, we integrate up to a maximum separation of $r=200$ \mpch\ using the measured $\xi(r)$ up to $r=30$ \mpch, and extrapolating it to $r=200$ \mpch\- assuming a linear theory dark matter correlation function with linear bias.

\begin{figure}
\begin{center}
\includegraphics[width=0.48\textwidth]{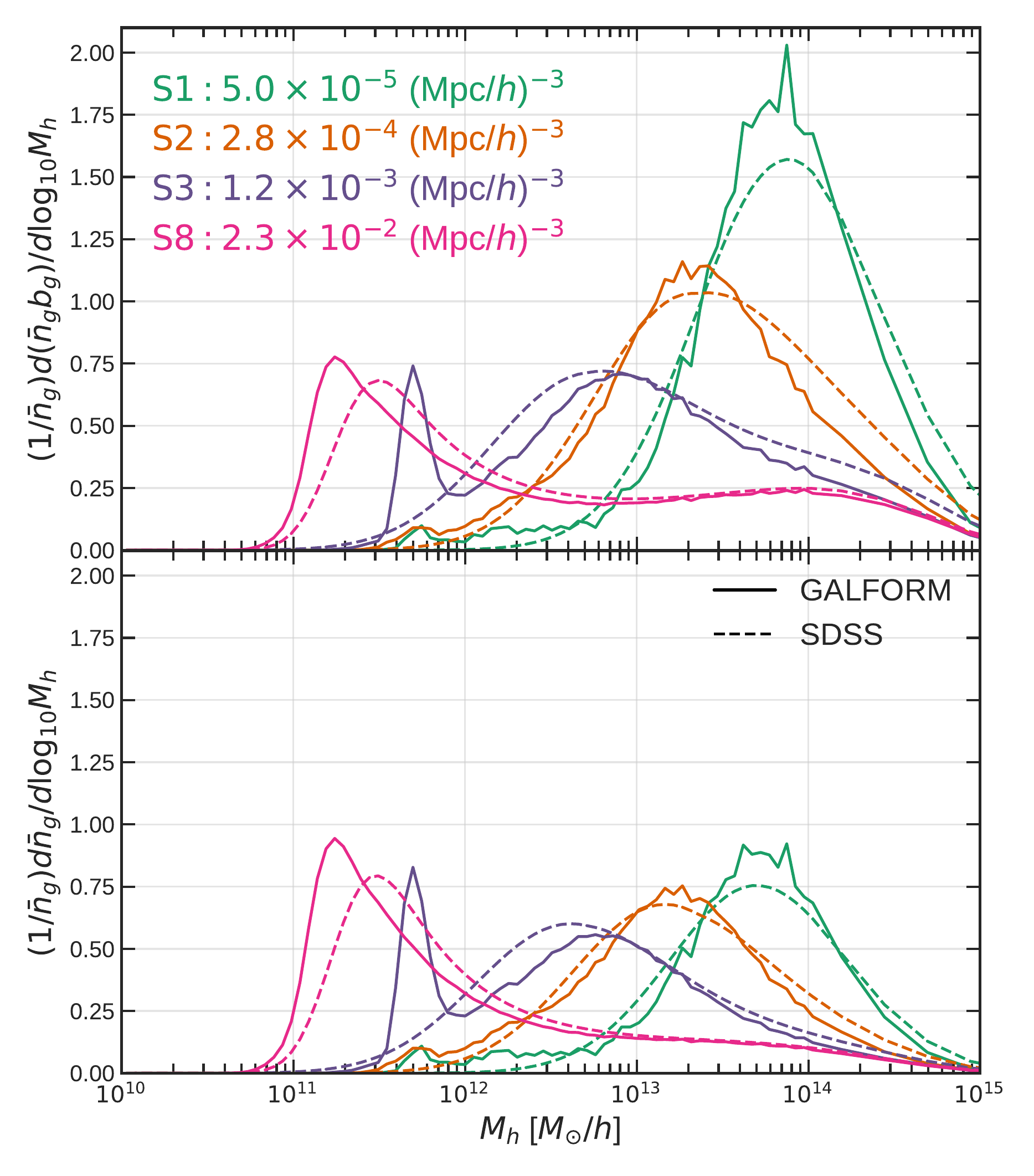}
\caption{Top panel: the integrand of the bias (Eq.~\ref{eq:biasint}) as a function of halo mass in the \galform\ samples (solid lines) and SDSS samples (dashed lines). The area under each curve is the large-scale bias of the sample. Bottom panel: integrand of the number density (Eq.~\ref{eq:numberdensint}) as a function of halo mass, normalised by the number density of each sample. The area under each curve is 1. Although the large-scale bias and number densities are the same for the \galform\ and SDSS samples, they differ in the halo masses that contribute in each sample.}
\label{fig:integrandSDSS}
\end{center}
\end{figure}

\begin{figure}
\centering
\subfloat{
\includegraphics[width=0.48\textwidth]{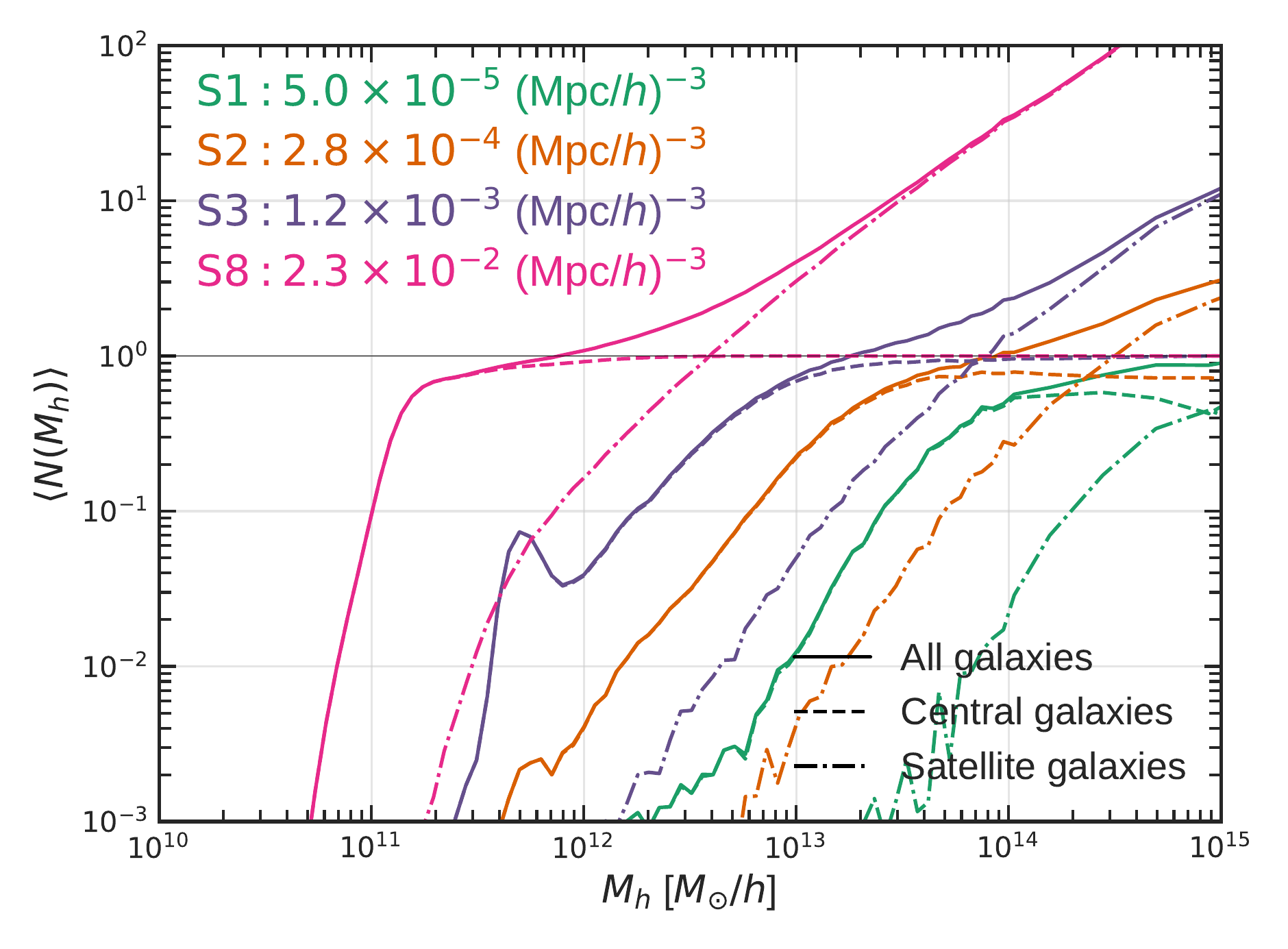}
}
\\
\subfloat {
\includegraphics[width=0.48\textwidth]{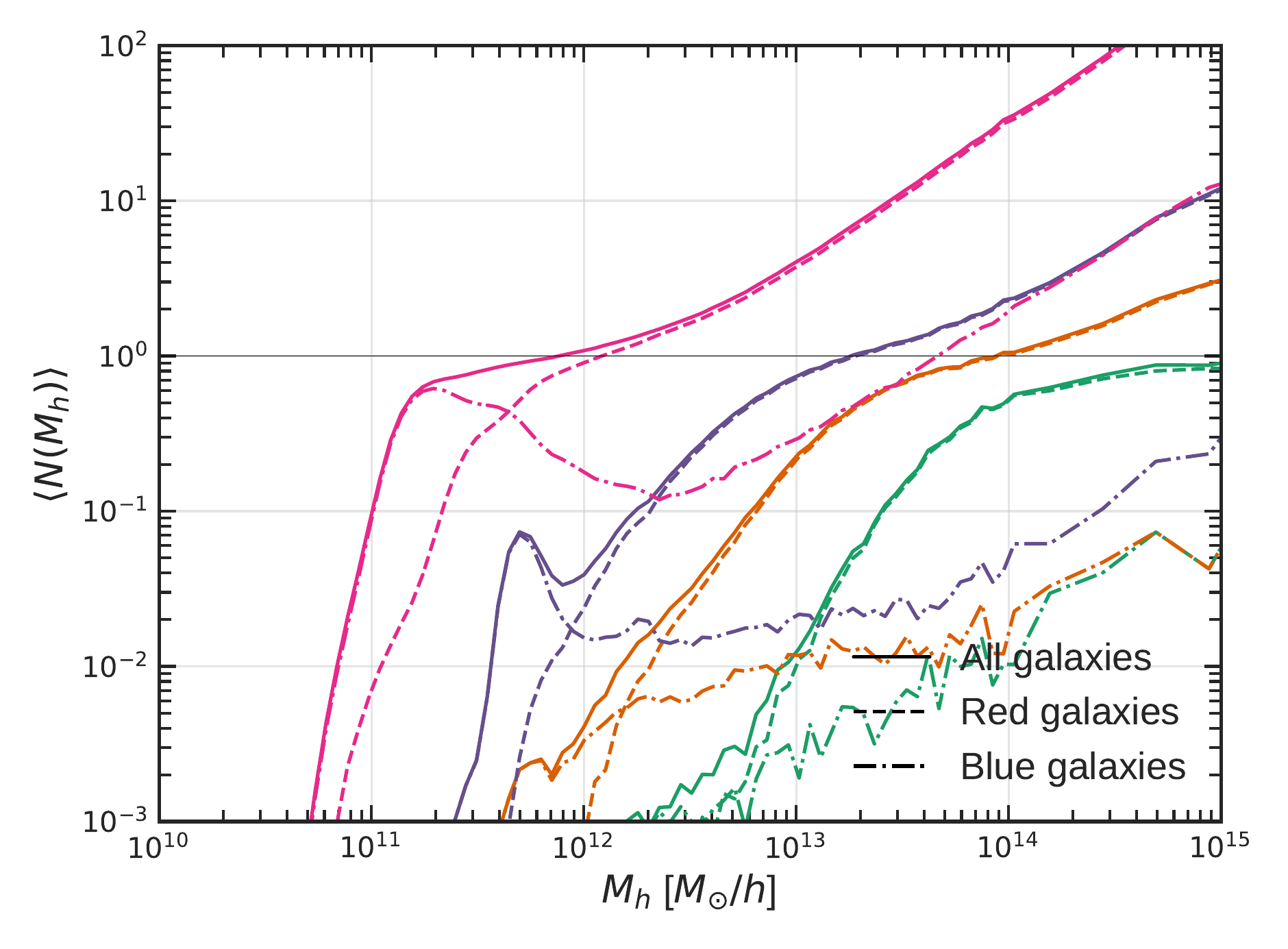}
}
\caption{Top panel: HOD of the four \galform\ samples split by central and satellite galaxies. Bottom panel: same as above, but split by red and blue galaxies according to the colour cut shown in Fig.~\ref{fig:gal_col}. In the top panel, the HODs of central galaxies are shown in dashed lines in each sample and those of satellites are shown in dot-dashed lines. In the bottom panel, the HODs of red galaxies are shown in dashed lines in each sample and those of blue galaxies are shown in the dot-dashed lines. Many of the non-standard features appear to come from blue central galaxies.}
\label{fig:galformHODsplit}
\end{figure}

\begin{figure*}
\centering
\subfloat{
\includegraphics[width=0.48\textwidth]{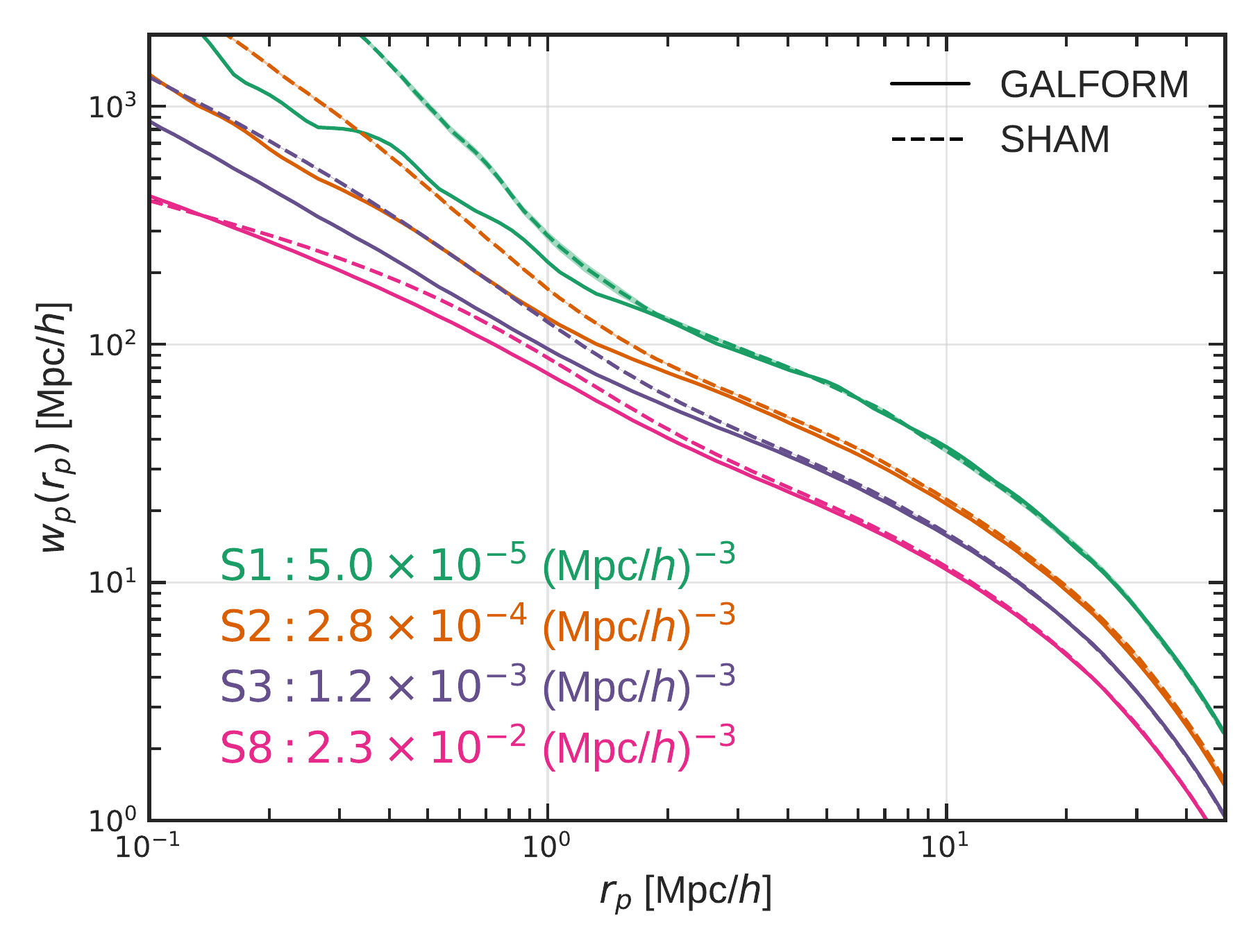}
}
\subfloat {
\includegraphics[width=0.48\textwidth]{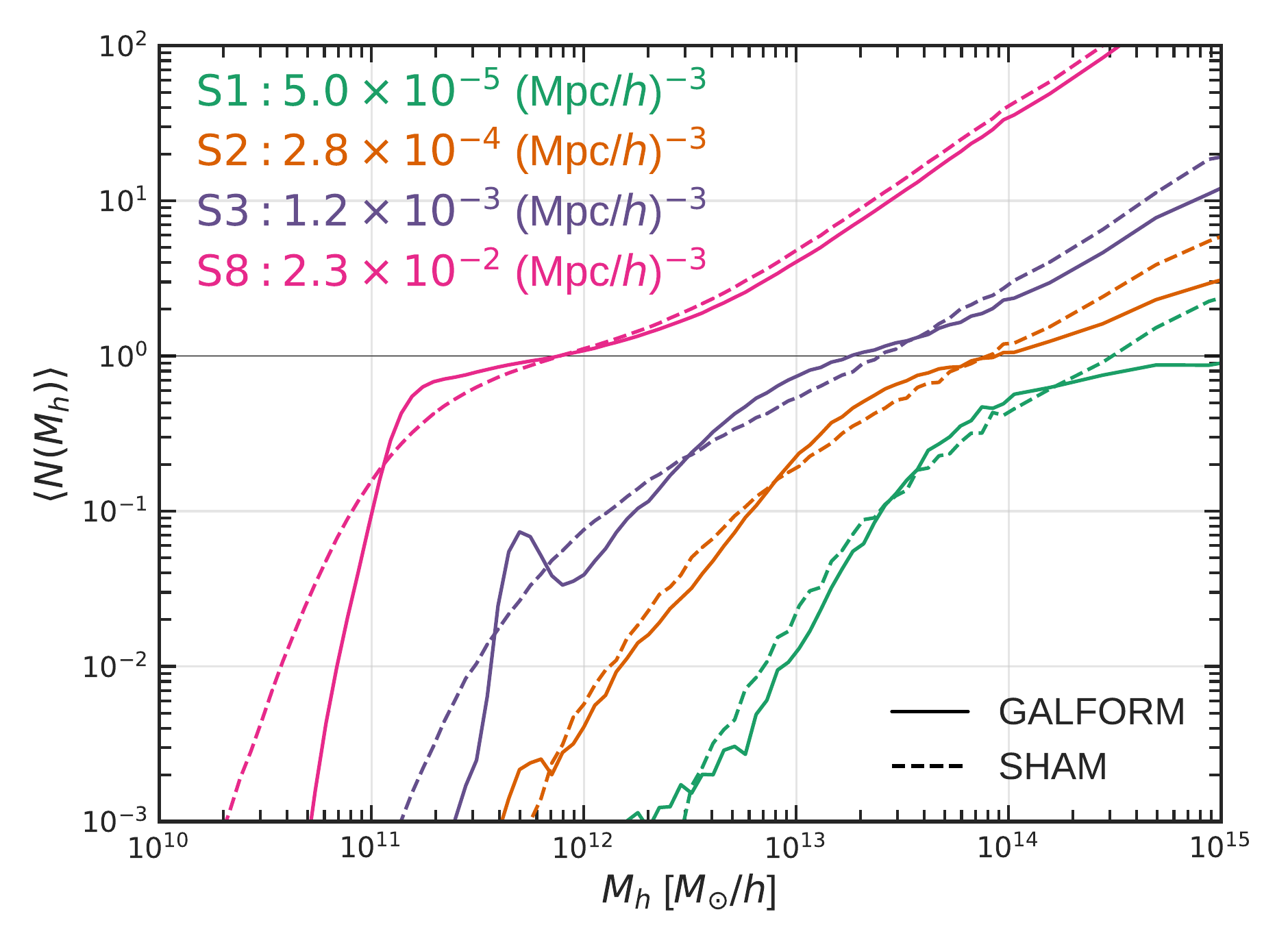}
}
\caption{Left panel: projected correlation function of \galform\ samples (solid lines) versus the average clustering from 10 SHAM catalogues (dashed lines). The shaded regions show the standard deviation in the clustering of the 10 SHAM catalogues. Right panel: the HODs of \galform\ (solid lines) and SHAM (dashed lines) samples, measured directly in all cases. The spread in HOD from the 10 SHAM catalogues is not shown because it is so small.}
\label{fig:galformSHAM}
\end{figure*}

\begin{figure}
\begin{center}
\includegraphics[width=0.48\textwidth]{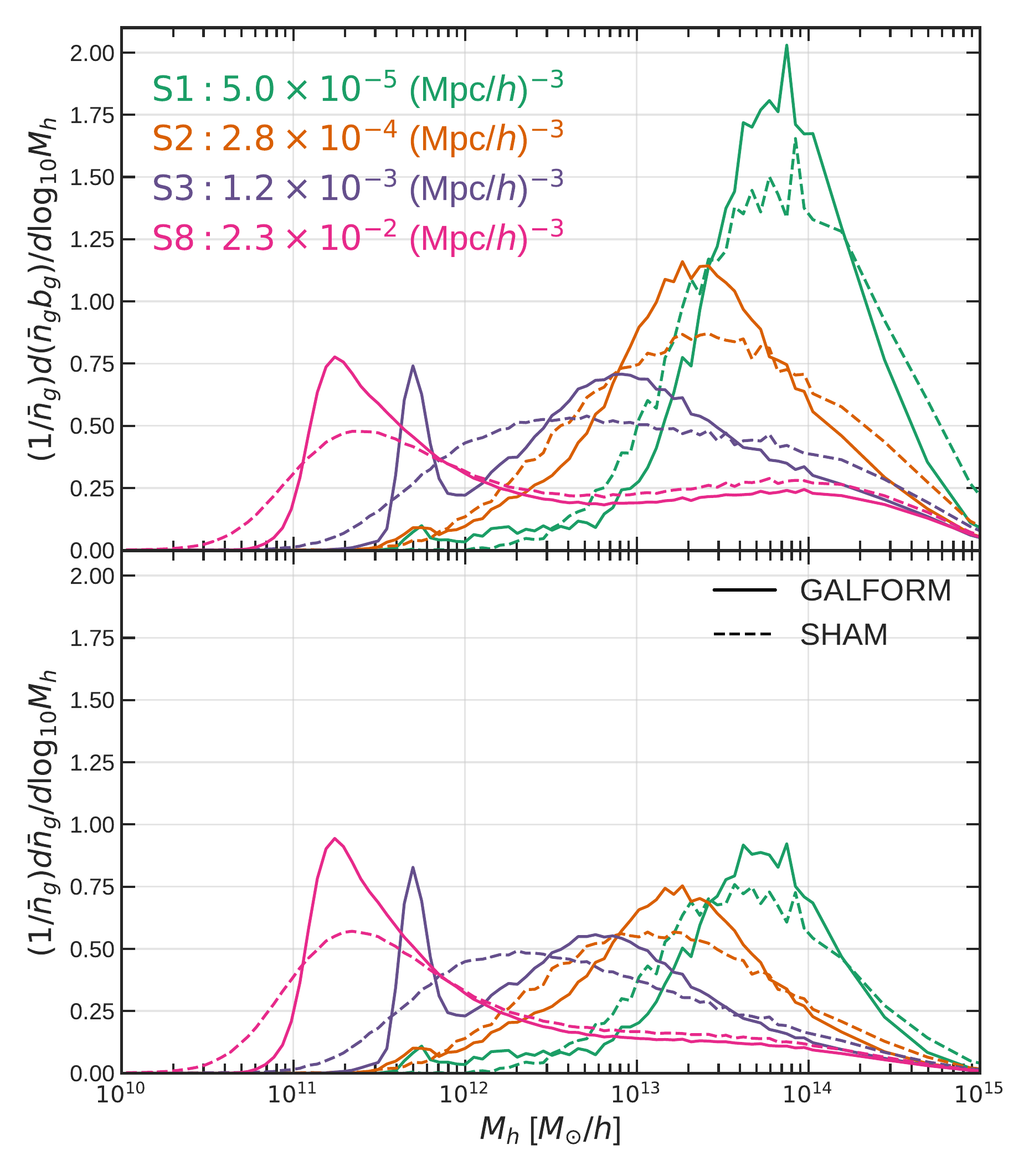}
\caption{Same as Fig. \ref{fig:integrandSDSS} but for \galform\ (solid lines) versus SHAM (dashed lines). The main differences between \galform\ and SDSS, shown in Fig.~\ref{fig:integrandSDSS}, seem to be preserved in the comparison between \galform\ and SHAM.}
\label{fig:integrandSHAM}
\end{center}
\end{figure}

In the left panel of Fig.~\ref{fig:galformSDSS}, we show the projected correlation functions of four \galform\ samples (S1, S2, S3, and S8 in Table~\ref{tab:samples}), along with those of the corresponding samples from SDSS \citep{zehavi2011}. The number densities of the samples are given in the figure legend. We highlight these samples in particular because they show a large range in number density and they are not very affected by the Sloan Great Wall \citep[unlike S4 and S5, see e.g.][for further details]{zehavi2011}. They provide a good match to the SDSS clustering, despite the model not being tuned to reproduce the SDSS clustering. 

The corresponding HODs are shown in the right panel of Fig.~\ref{fig:galformSDSS}. For the \galform\ samples, these are measured directly from the simulation, whereas in SDSS the HODs shown are the best-fit 5-parameter model (Eq.~\ref{eq:hodmodel}) to the galaxy clustering measurements \citep{zehavi2011}. For the faintest sample (S8), there is a weak constraint on the $\sigma_{\log M}$ parameter from SDSS, so we also show an HOD with a low $\sigma_{\log M}$ imposed, to give a sense for the range of models that fit the data well. Although we do not fit HODs to the \galform\ clustering measurements, we can assume that they will have fairly similar shapes to the best fit HODs of SDSS, since the clustering signal in the five samples is similar.

There are several interesting features in the HODs of the \galform\ samples when compared to the best fit SDSS HOD models. The most glaring difference is the feature in the $\bar n=1.2\times 10^{-3}$ (Mpc/$h$)$^{-3}$ sample (S3) between $10^{11}<M_h<10^{12}$\msolarh , which is highlighted with a shaded box in the figure. The feature is also present to some extent in the $\bar n=2.8\times 10^{-4}$ (Mpc/$h$)$^{-3}$ sample (S2) at the same mass. Such a feature will never be captured by a standard parameterization of the HOD, such as the 5-parameter model we are considering, because they are smooth and monotonic by construction. 

An interesting aspect of this feature is that it appears at $\langle N \rangle < 0.1$, while most HOD model fits are presented for $\langle N \rangle > 0.1$. One might wonder how significant such a feature is to the overall bias and number density of the sample. We can answer that by looking at the integrands in the Eqs.~\ref{eq:numberdensint} and~\ref{eq:biasint}.

The top panel of Fig.~\ref{fig:integrandSDSS} shows the HODs in Fig.~\ref{fig:galformSDSS} multiplied by the \citet{jenkins2001} mass function, which is the quantity inside the integral in Eq.~\ref{eq:biasint}. The area below each curve gives the bias of that sample, and these values are given in Table~\ref{tab:samples} ($b_g$ from HOD). We find that the values of the bias measured from the HODs are significantly different from those measured from the 2-point clustering for the faintest samples (S3 and S8). We attribute this to assembly bias, which we expect to be present in \galform\ because it is built from subhalo merger trees, and by construction is absent in the standard HOD formalism. As we discuss in \S\ref{sec:discussion}, if we remove assembly bias by shuffling galaxies within a halo mass bin, the bias from the clustering of these shuffled catalogues agrees within the error for each sample with that measured from the HODs. 

The bottom panel of Fig.~\ref{fig:integrandSDSS} shows the integrand of the number density of the sample, which is the quantity inside the integral in Eq.~\ref{eq:numberdensint}, normalized by the number density of the given sample (so the area under each curve is unity). These two figures show how much various features in the HODs contribute to the overall number density and bias of the sample. So although the feature in the HOD of S3 is below $\langle N \rangle = 0.1$, it contributes significantly (roughly 10\%) to the bias and number density of that sample. This is due to the fact that the HOD is multiplied by the halo mass function in Eqs.~\ref{eq:numberdensint} and~\ref{eq:biasint}, and the halo mass function is steep, so low halo masses get a relatively high weight.

Another notable difference in the HOD shapes appears in the faintest sample (S8), where we see in Fig. \ref{fig:galformSDSS} that the turn-over that is imposed to be at $\langle N \rangle=1$ in standard HOD models is at roughly 0.8 occupation in the \galform\ sample. Regardless of the value of $\sigma_{\log M}$, the commonly used 5-parameter HOD model cannot perfectly fit this feature, because it assumes a step function at $\langle N \rangle=1$. 

In all of the \galform\ HODs, the satellite contribution lies below that predicted from SDSS. In the brightest sample (S1), the occupation barely reaches $\langle N \rangle=1$ at the highest halo masses shown here ($M_h=10^{15}$ \msolarh). Note that the clustering of \galform\ deviates from that of SDSS on small scales (1-halo term) in the brighter samples, which means we should not expect the HODs to be exactly equivalent, especially for the satellites.

\begin{table*}
\centering
\caption{Luminosity-threshold samples used in SDSS clustering and HOD analysis as defined in \citet{zehavi2011}. Columns are: number density, SDSS luminosity threshold, the corresponding \galform\ luminosity threshold, fraction of orphan galaxies, the amount of scatter needed in our SHAM method to match the large-scale clustering amplitude of the \galform\ sample, the large-scale bias computed by multiplying the measured HOD by the mass function and integrating over $M_h$ (Eq.~\ref{eq:biasint}), and the large-scale bias estimated from the 2-point correlation function between 10 and 40 \mpch. The difference in these two bias values quantifies the level of assembly bias in the sample, as the bias from the HOD does not include assembly bias, whereas the 2-point correlation function does. We only make SHAM catalogues for S1, S2, S3, and S8.}
\label{tab:samples}
\begin{tabular}{ccccccccc}
Sample&$\bar n$ (Mpc/$h$)$^{-3}$ &SDSS $M_r^{\textrm{faint}}$  & \galform\ $M_r^{\textrm{faint}}$&$f^{\textrm{orphan}}$&SHAM $\Delta M_r$&$b_g$ from HOD&$b_g$ from $\xi(r)$\\
\hline 
S1& $5.0\times10^{-5}$ &-22.0 & -22.50 & 0.0038 & 0.6&1.98&2.03\\
S2& $2.8\times10^{-4}$&-21.5  & -21.88 & 0.0050 & 0.8&1.57&1.55\\
S3& $1.16\times10^{-3}$&-21.0  & -21.12 & 0.0098 & 1.1&1.31&1.34\\
S4& $3.18\times10^{-3}$ & -20.5 & -20.54 &0.0118&&1.15&1.16\\
S5&  $6.56\times10^{-3}$ & -20.0 & -20.06 & 0.0168&&1.10&1.13\\
S6&$1.120 \times 10^{-2}$ & -19.5 & -19.57 & 0.0267 &&1.10&1.14\\
S7&  $1.676\times10^{-2}$ & -19.0& -19.10 & 0.0380 &&1.09&1.14\\
S8& $2.311\times10^{-2}$&-18.5  & -18.66 &0.0486 & 1.2&1.09&1.13\\
S9 &  $3.030\times10^{-2}$& -18.0 & -18.26 &0.0591 &&1.08&1.13
\end{tabular}
\end{table*}

\begin{table*}
\centering
\caption{Parameters of best fit 5-parameter HOD models in each of the four samples considered, showing in turn for each sample the results from SDSS \citep[from][]{zehavi2011}, \galform\ and SHAM. For the \galform\ and SHAM samples, the mean of 50 bootstrap realizations is given, and the errors quoted are the standard deviations of these realizations. As mentioned in the text, the halo masses of \galform\ and SHAM have been converted to the equivalent of $M_{m,200}$ in a WMAP7 cosmology.}
\label{tab:bestfit}
\begin{tabular}{c c | c | c c c c c }
&&&\multicolumn{2}{|c|}{Central} &\multicolumn{3}{|c|}{Satellite} \\
Sample &$\bar n$ (Mpc/$h$)$^{-3}$ & & $\log M_{\textrm{min}}$ & $\sigma_{\log M}$ & $\log M_0$ & $\log M_{1}'$ & $\alpha$  \\
\hline \hline
S1 &$5.0\times10^{-5}$ & SDSS   & 14.06 $\pm$ 0.06 & 0.71 $\pm$ 0.07 & 13.72 $\pm$ 0.53& 14.80 $\pm$ 0.08    & 1.35 $\pm$ 0.49 \\
			 &     & \galform\ & 13.94 $\pm$ 0.01 & 0.59 $\pm$ 0.01 & 10.34 $\pm$ 0.65 & 17.48 $\pm$ 0.42 & 1.81 $\pm$ 0.20 \\
			  &    & SHAM   & 14.07 $\pm$ 0.01 & 0.75 $\pm$ 0.01 & 13.77 $\pm$ 0.38 & 14.78 $\pm$ 0.12 & 1.22 $\pm$ 0.37 \\
\hline
S2 &$2.8\times10^{-4}$ & SDSS     & 13.38 $\pm$ 0.07 & 0.69 $\pm$ 0.08 & 13.35 $\pm$ 0.21 & 14.20 $\pm$ 0.07 & 1.09 $\pm$ 0.17  \\
			&     & \galform\ & 13.60 $\pm$ 0.07 & 0.85 $\pm$ 0.05 & 12.72 $\pm$ 0.09 & 14.69 $\pm$ 0.18 & 0.30 $\pm$ 0.19 \\
			&     & SHAM     & 13.52 $\pm$ 0.01 & 0.85 $\pm$ 0.01 & 13.22 $\pm$ 0.10 & 14.23 $\pm$ 0.02 & 1.02 $\pm$ 0.10  \\
\hline
S3& $1.2\times10^{-3}$ & SDSS    & 12.78 $\pm$ 0.10 & 0.68 $\pm$ 0.15 & 12.71 $\pm$ 0.26 & 13.76 $\pm$ 0.05 & 1.15 $\pm$ 0.03  \\
			&     & \galform\ & 12.85 $\pm$ 0.02 & 0.77 $\pm$ 0.02 & 12.37 $\pm$ 0.05 & 13.83 $\pm$ 0.02 & 0.97 $\pm$ 0.06 \\
			&     & SHAM    & 12.91 $\pm$ 0.01 & 0.85 $\pm$ 0.01 & 13.15 $\pm$ 0.10 & 13.65 $\pm$ 0.02 & 0.99 $\pm$ 0.03 \\
\hline
S8& $2.3\times10^{-2}$ & SDSS & 11.33 $\pm$ 0.07 & 0.26 $\pm$ 0.21 & 8.99 $\pm$ 1.33 & 12.5 $\pm$ 0.04 & 1.02 $\pm$ 0.03\\
			&      & \galform\ & 11.25 $\pm$ 0.01 & 0.25 $\pm$ 0.01 & 9.91 $\pm$ 1.40 & 12.59 $\pm$ 0.08 & 1.07 $\pm$ 0.06 \\
			 &     & SHAM & 11.41 $\pm$ 0.01 & 0.51 $\pm$ 0.01 & 9.02 $\pm$ 0.01 & 12.51 $\pm$ 0.01 & 1.06 $\pm$ 0.01 
\end{tabular}
\end{table*}

In order to further investigate the non-standard features seen in the \galform\ HOD, we split our galaxy samples in a number of ways. Fig.~\ref{fig:galformHODsplit} shows the \galform\ HODs for the four samples, split into central and satellite galaxies (top panel), and red and blue galaxies (bottom panel). The red-blue split used is given by the black line in Fig.~\ref{fig:gal_col}. As can be seen in this figure, the galaxies that contribute the bump to S3 are largely blue central galaxies. These blue centrals also contribute to the shape of the turn-over in S8 \citep{contreras2015, gonzalez2017}.

There are several notable features in Fig.~\ref{fig:galformHODsplit} that help explain the non-standard shapes of the \galform\ HODs. In the brightest two samples, $\bar n= 2.3 \times 10^{-2}$ and $1.2\times 10^{-3}$ (Mpc/$h$)$^{-3}$ (S1 and S2), the mean occupation of central galaxies does not plateau at $\langle N \rangle = 1$ as is assumed in the standard HOD formalism. Also, the occupation of blue galaxies in \galform\ is not monotonic for the faintest two samples, $\bar n= 2.8 \times 10^{-4}$ and $5.0\times 10^{-5}$ (Mpc/$h$)$^{-3}$ (S3 and S8). These features indicate that the \galform\ HODs cannot be well described within the standard HOD formalism.

We attribute the feature at $M_h \approx 5\times 10^{11}$ \msolarh in the HOD of S3 to AGN feedback. \galform\ assumes an equation in which AGN energy prevents the gas from cooling in haloes in which the dynamical time is less than the cooling time. This results in a transition in behaviour for haloes of around $5\times10^{11}$\msolarh~\citep{bower2006}. For a fixed stellar mass, AGN feedback decreases the luminosity of galaxies by quenching star formation, so formerly blue galaxies become redder. While the relationship between halo mass and stellar mass may be monotonic, the relationship between halo mass and luminosity may not be when AGN feedback is included. We also looked at the HODs in \galform\ with AGN feedback switched off, and we found the sharp feature in S3 was not present, and the HODs had more standard shapes. We do not show this model here because it was not tuned to match observational constraints, and thus cannot be compared on equal footing.

While AGN feedback may be the main culprit for the particular feature in S3, there are other factors that are important in shaping the HOD in \galform, such as disk instabilities and variations in the strength of supernova feedback \citep{mitchell2016}.

\begin{figure}
\centering
\subfloat{
\includegraphics[width=0.48\textwidth]{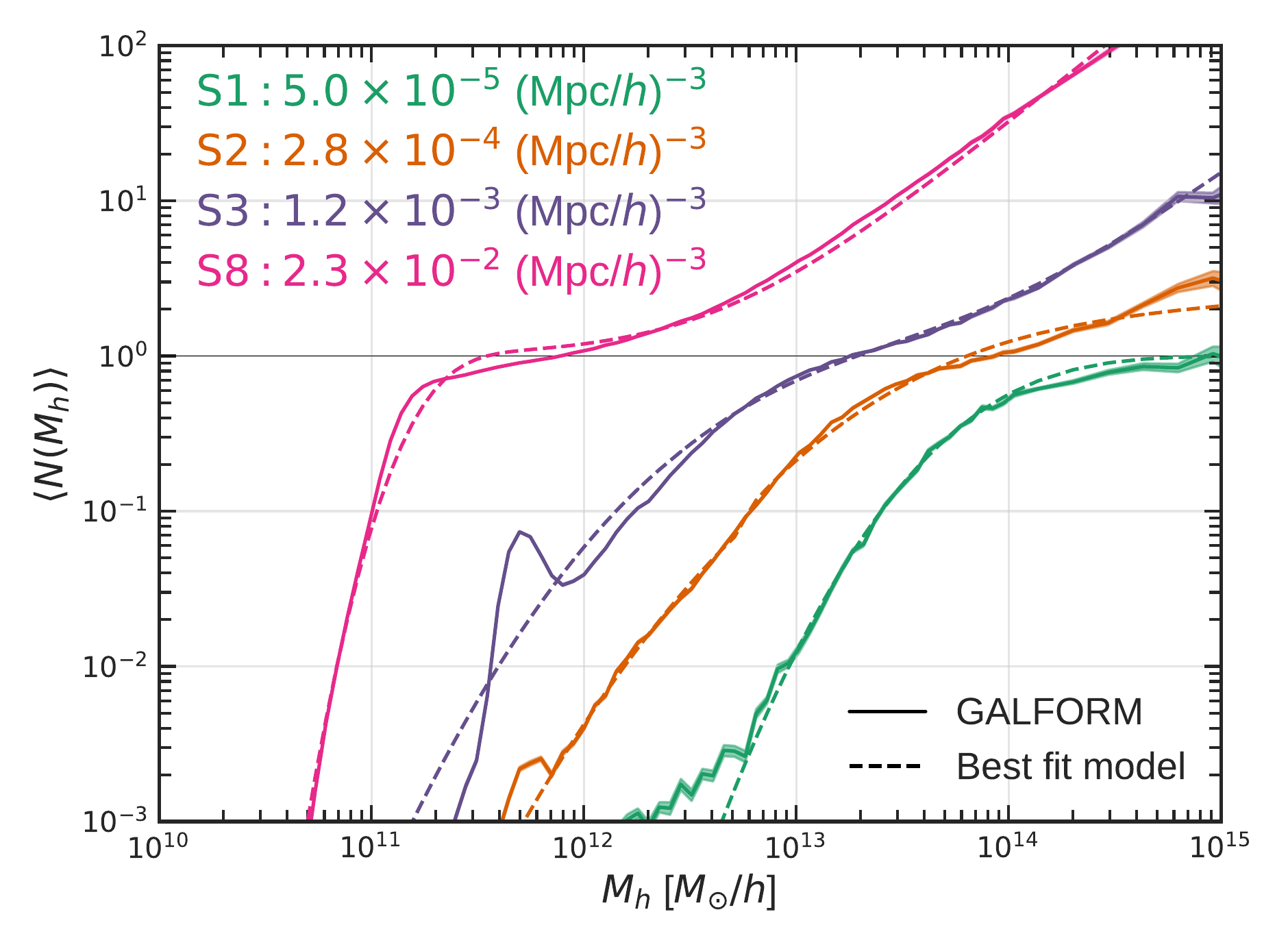}
}
\\
\subfloat {
\includegraphics[width=0.48\textwidth]{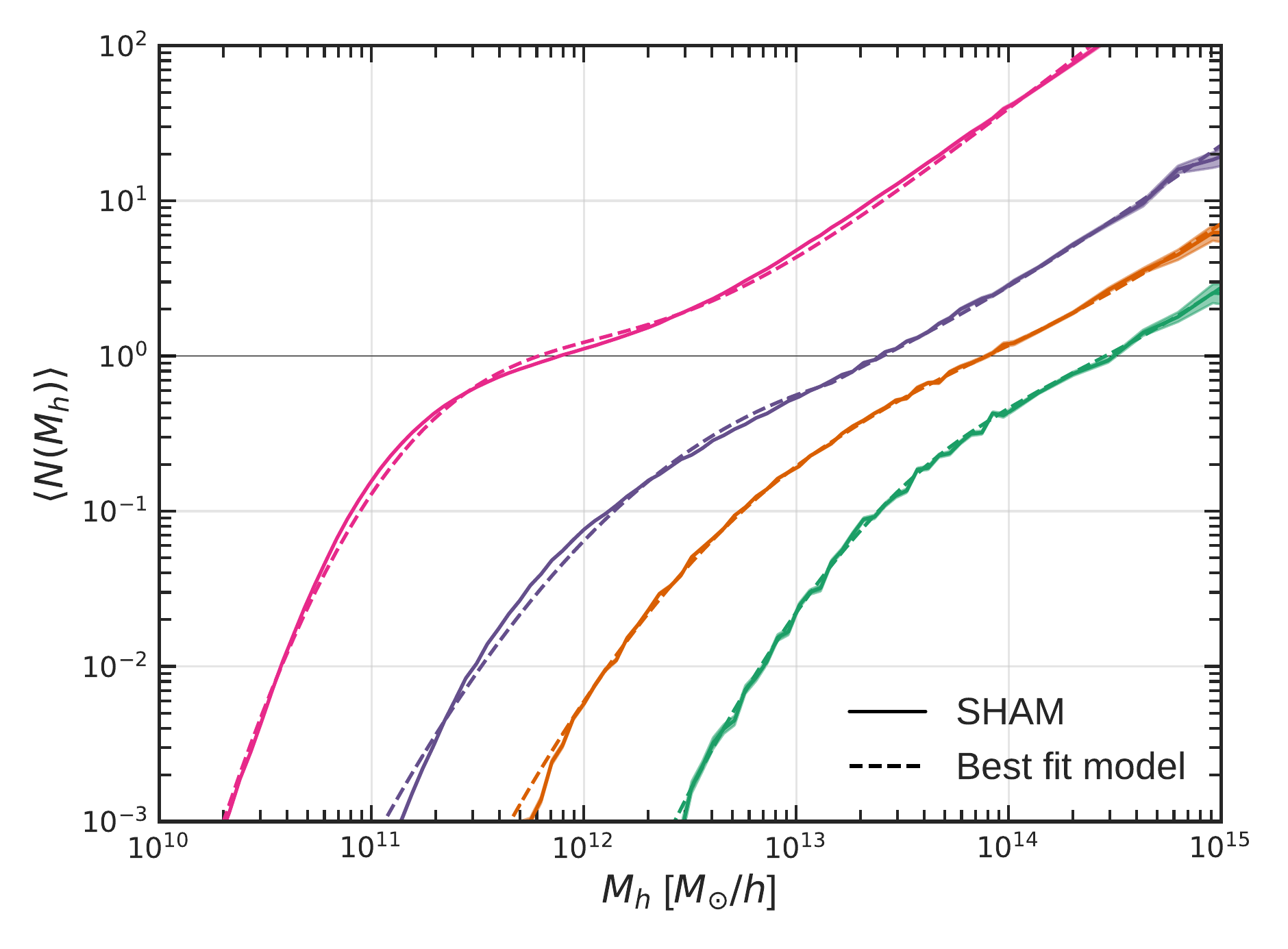}
}
\caption{Top panel: Solid lines and shaded regions show the mean and 1-sigma spread in the HODs measured from 50 bootstrap samples from \galform. Dashed lines show the best-fitting 5-parameter model to the solid lines of the same colour. Bottom panel: same as above but for SHAM samples. Dashed lines again show the best-fitting 5-parameter model to the solid lines of the same colour. SHAM HODs are better described by standard HOD model than \galform\ HODs.}
\label{fig:bestfit}
\end{figure}

\subsection{Contrasting GALFORM and SHAM}

\label{sec:galsham}

We create 10 SHAM catalogues from each of the \galform\ samples using the method described in \S\ref{sec:SHAM}. To tune the scatter, We try a range of values ($0.5 \leq \Delta M_r \leq 1.5$ in steps of size 0.1), measure the average clustering in the resulting SHAM samples, and choose the value of the scatter that best matches the large-scale clustering of the given \galform\ sample. The values of $\Delta M_r$ needed for the four samples are given in Table~\ref{tab:samples}. The average projected correlation function of these SHAM samples is shown with dashed lines in the left panel of Fig.~\ref{fig:galformSHAM}, along with those of the \galform\ samples in solid lines. The shaded regions show the 1-sigma spread in the 10 SHAM realisations of each sample. 

The HODs of the four samples are shown in the right panel of Fig.~\ref{fig:galformSHAM}. These HODs are measured directly for both the \galform\ and SHAM samples, as we have access to all of the information needed to measure the HOD. Here we show the HODs from a single SHAM realisation, as there is almost no spread in the HOD shape between the 10 realisations. As mentioned previously, all halo masses shown have been converted to the equivalent of $M_{m,200}$ in a WMAP7 cosmology.

The satellite contributions in the SHAM HODs are generally steeper than those in \galform, which we might expect due to the difference in small-scale clustering between the samples. We also see that the features in the \galform\ HODs that we mentioned in \S\ref{sec:galformSDSS} are not preserved when SHAM is applied. In particular, the feature in S3 in \galform\ is not present in the corresponding SHAM sample. 

Overall, the SHAM HODs have a more standard (monotonic) shape. The bias and number density integrands are shown in Fig.~\ref{fig:integrandSHAM}. These show similar differences to what is seen in the comparison to the SDSS best-fit HODs. For clarity, we show only the integrand for a single SHAM realisation in Fig.~\ref{fig:integrandSHAM}, because the spread between the SHAM realisations is very small.

Next we examine how well the 5-parameter HOD model describes the HODs measured in \galform\ and SHAM. We quantify this by doing a least-squares fit of the model to the measured HODs in each sample. In order to estimate the errors on the 5 parameters, we fit the model to 50 bootstrap realizations (with replacement) of the haloes in each galaxy sample and look at the spread in the best-fitting parameter values. The fits are done in $\log_{10} M_h$ and $\log_{10} \langle N(M_h) \rangle$, in the range $M_h < 10^{14.9}$ \msolarh, $\langle N(M_h)\rangle > (10^{-2.0}, 10^{-2.5}, 10^{-3.0}, 10^{-3.0})$ for (S1, S2, S3, S8), respectively, and with the constraint that the number density of the sample is fixed. We also assume uniform errors as a function of halo mass. Table~\ref{tab:bestfit} gives the mean and standard deviation of the best-fitting parameters of the bootstrap realizations of the \galform\ and SHAM samples, along with the best-fitting parameters from the SDSS clustering from \citet{zehavi2011}. 

As can be seen from Table~\ref{tab:bestfit}, the brightest sample (S1) in \galform\ gives very poor constraints on the parameters associated with satellite occupation ($\log M_0$, $\log M_1'$, and $\alpha$). This is because the HOD of that sample only barely reaches $\langle N \rangle =1$ in the range of halo masses considered.

Fig.~\ref{fig:bestfit} shows the measured HOD and best fit HODs for \galform\ (top panel) and SHAM (bottom panel) samples. Note that the best fit HOD models shown use the best-fitting parameters to the actual measured HOD, as opposed to the mean of the bootstrap realizations, but the values are very close in all cases. It is clear from this figure that the SHAM samples have HODs that are better described by the 5-parameter HOD model than the  HODs of the \galform\ samples. 

One way to quantify how well the model fits the data is through the sum of the squares of the residuals: $S\equiv\sum \big[\log_{10} \langle N(M_h) \rangle^{\rm data} - \log_{10} \langle N(M_h) \rangle^{\rm model}\big]^2$. For S3 in \galform, for example, $S=2.2$, whereas for SHAM $S=0.22$. This quantity is lower in SHAM for all samples, indicating that the model is a better fit to the SHAM HODs than it is to the \galform\ HODs. From Fig.~\ref{fig:bestfit} it is clear that the 5-parameter model does not provide a good fit to the curvature of the \galform\ HODs for the brightest two samples, and as was mentioned previously, the AGN feature in S3 and the turn-over at $\langle N\rangle<1$ in S8 are not well fit by the model. However, the clustering on large scales of SHAM and \galform\ is well matched in all cases. This highlights one of the limitations of HOD modeling and the information that can be inferred from it.

\begin{figure}
\centering
\subfloat{
\includegraphics[width=0.48\textwidth]{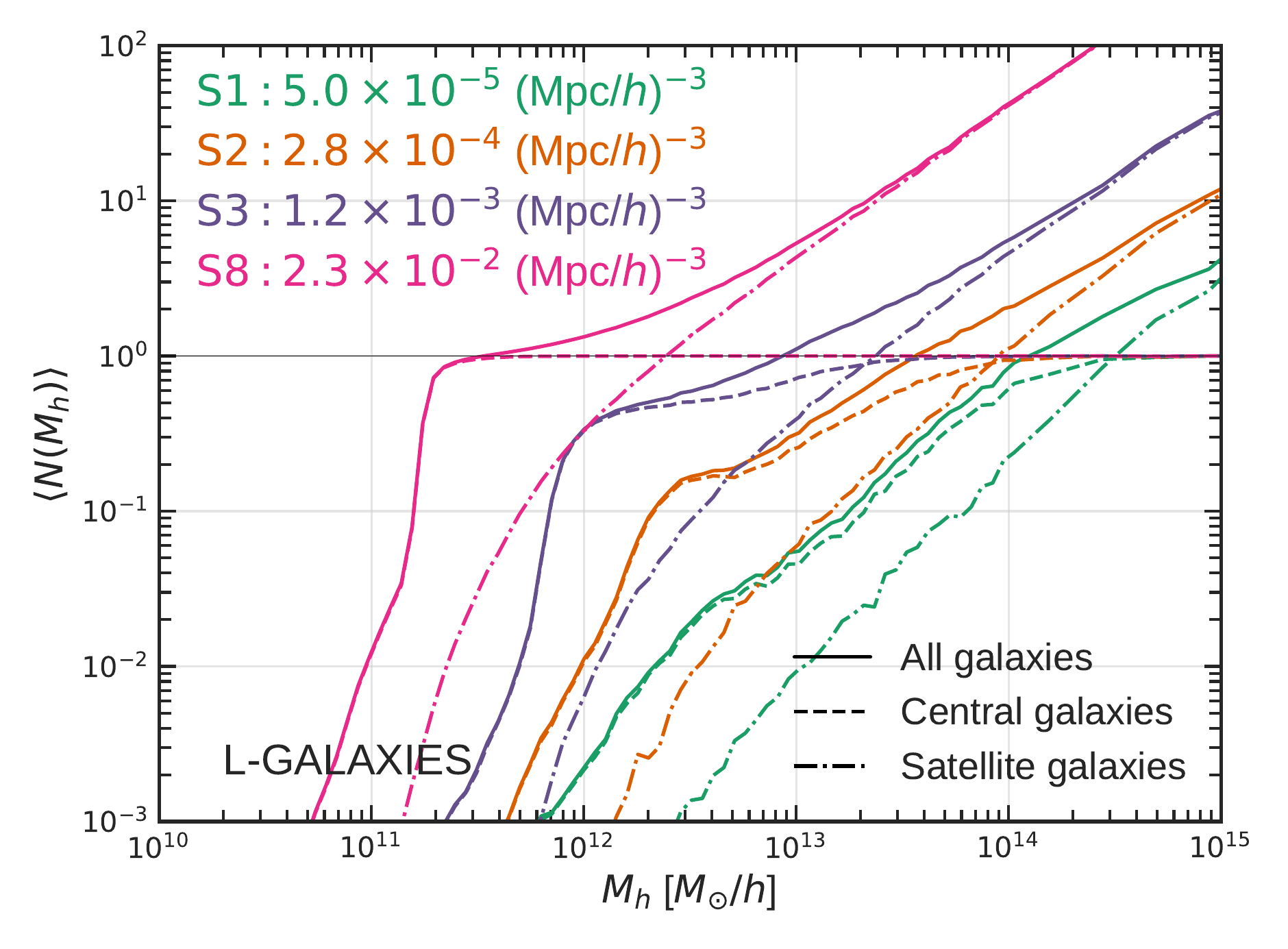}
}
\\
\subfloat {
\includegraphics[width=0.48\textwidth]{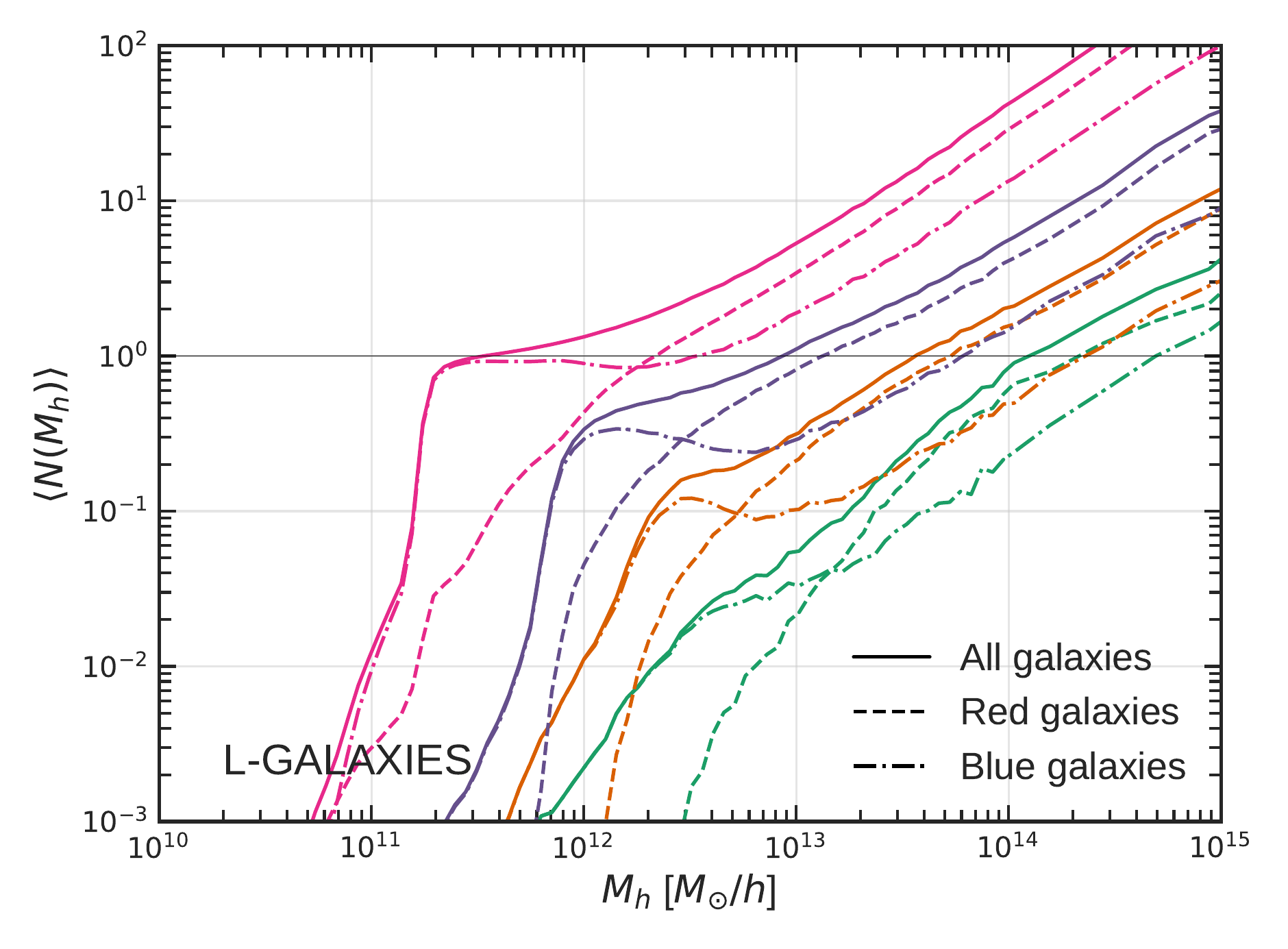}
}
\caption{HODs of galaxy samples from \lgalaxies\ (like Fig. \ref{fig:galformHODsplit}). Top panel: HODs split by centrals (dashed lines) and satellites (dot-dashed lines). Bottom panel: HODs split by red (dashed lines) and blue (dot-dashed lines) galaxies. Solid lines show HODs of all galaxies in both panels. As in \galform\, there are features that deviate from the standard HOD form that arise largely from blue central galaxies.}
\label{fig:lgals_hod}
\end{figure}

\begin{figure}
\centering
\subfloat{
\includegraphics[width=0.48\textwidth]{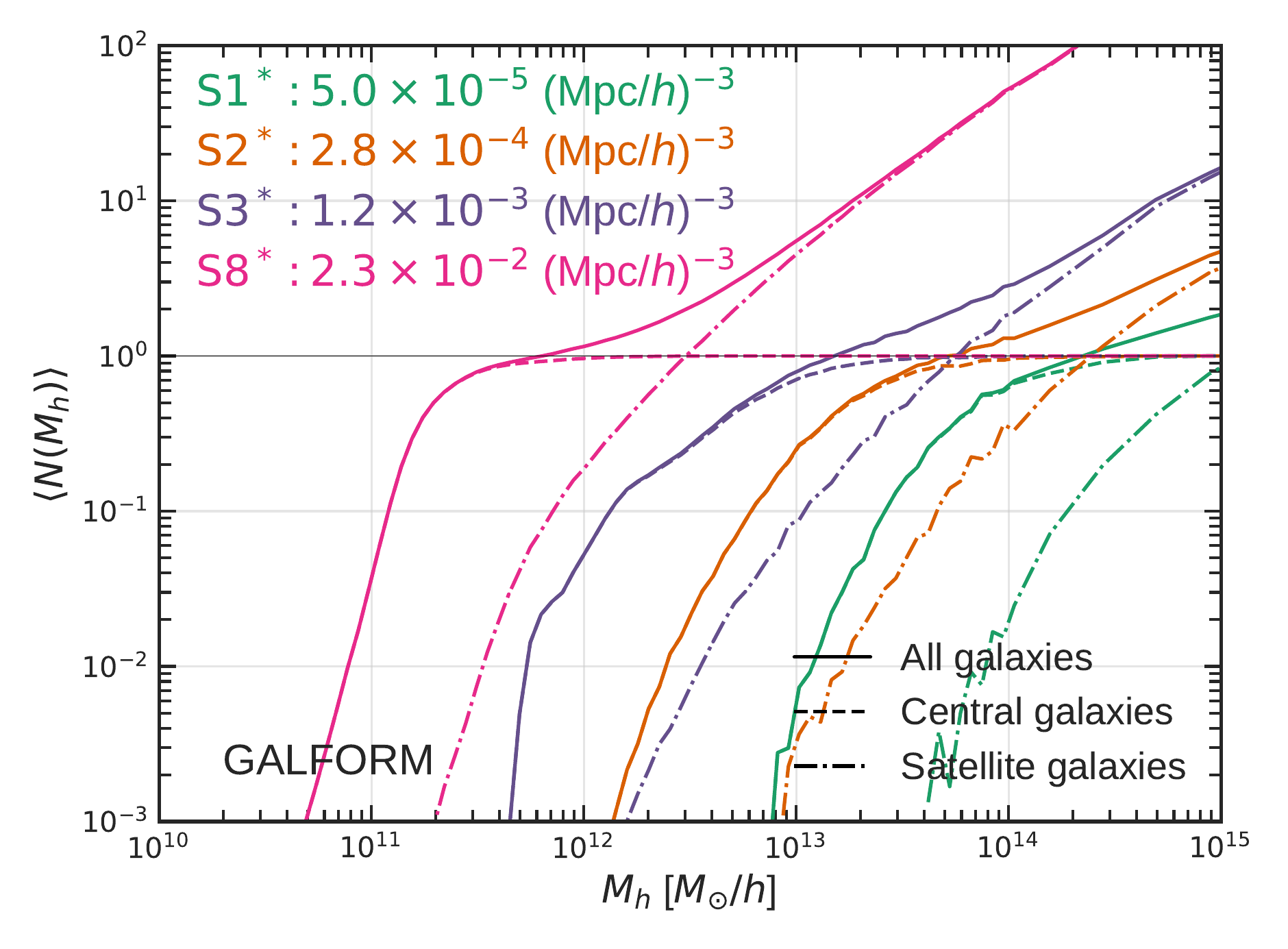}
}
\\
\subfloat {
\includegraphics[width=0.48\textwidth]{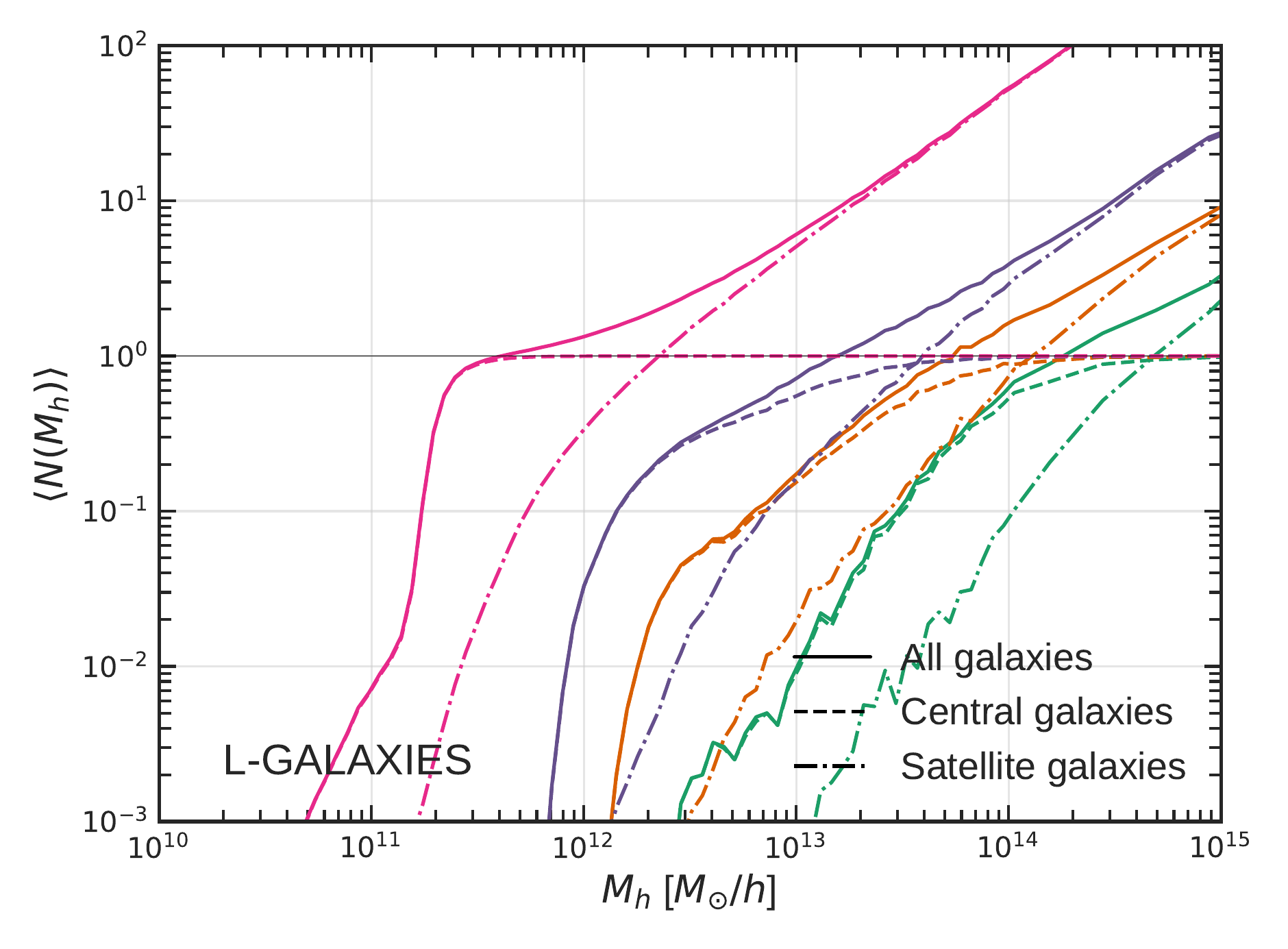}
}
\caption{Top panel: HODs of stellar-mass-threshold samples in \galform, split into centrals (dashed lines) and satellites (dot-dashed lines). Bottom panel: Same as above but for \lgalaxies\ stellar-mass-threshold samples. These HODs have more standard shapes, indicating that stellar mass is a more suitable property for HOD modelling.}
\label{fig:galformlgalsmstar}
\end{figure}

\section{Discussion}
\label{sec:discussion}

\subsection{Shape of the HOD}

\label{sec:discussion1}
In order to discuss our results in a broader context, it is important to understand how generic the HODs in \galform\ are: are these HODs non-standard because of the specific way AGN feedback is implemented in the model, or do other semi-analytic models that include AGN feedback show similar non-standard HODs? To address this question, we look at the HODs of luminosity-threshold samples in another semi-analytic model, \lgalaxies\ \citep{guo2011, guo2013}. AGN feedback in \lgalaxies\ is implemented in a very different way than in \galform: the feedback is proportional to black hole accretion rate which is based on Bondi accretion arguments \citep{croton2006}. Unlike in \galform, AGN feedback in \lgalaxies\ does not have a specific halo mass scale. 

\begin{figure*}
\centering
\subfloat{
\includegraphics[width=0.48\textwidth]{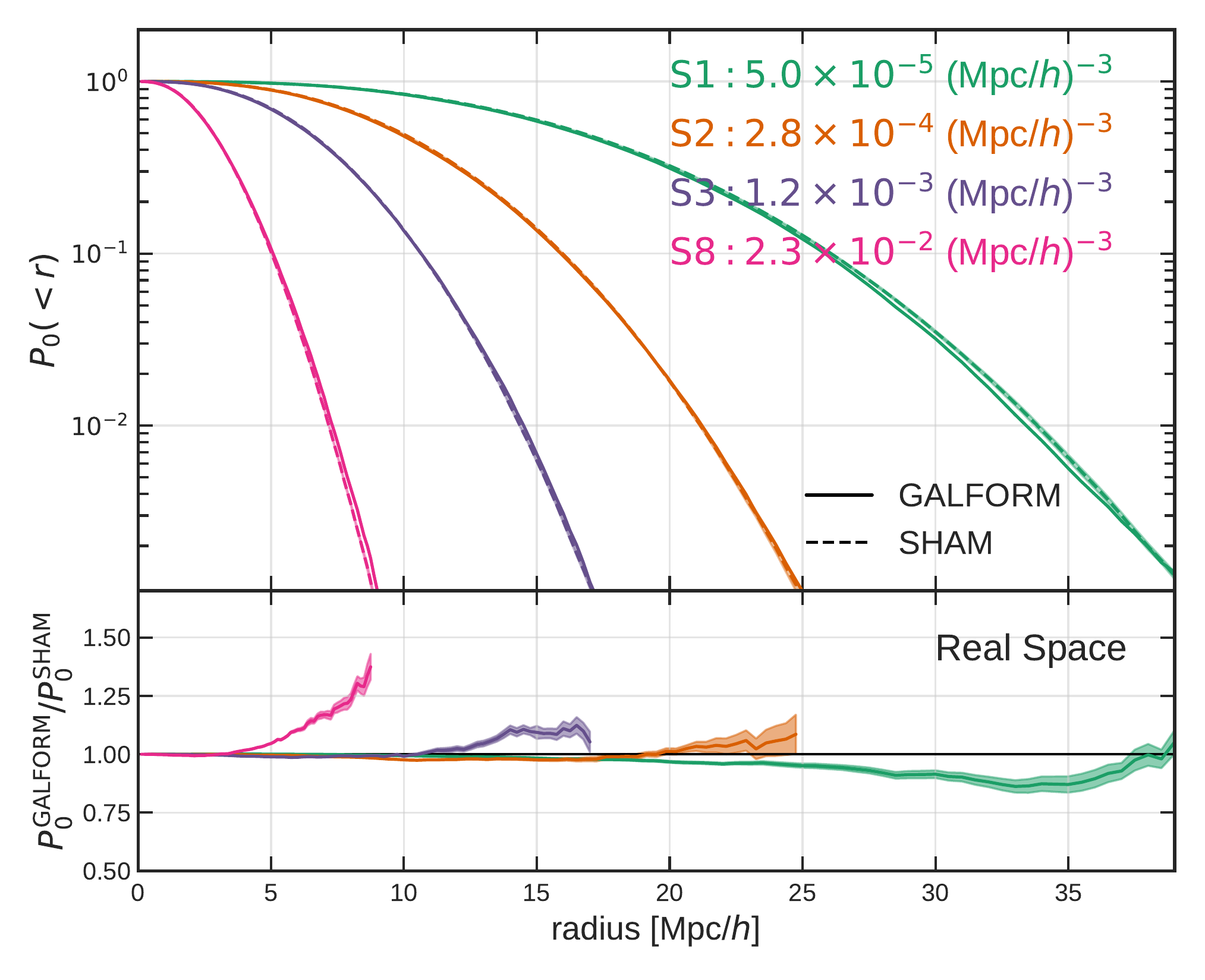}
}
\subfloat {
\includegraphics[width=0.48\textwidth]{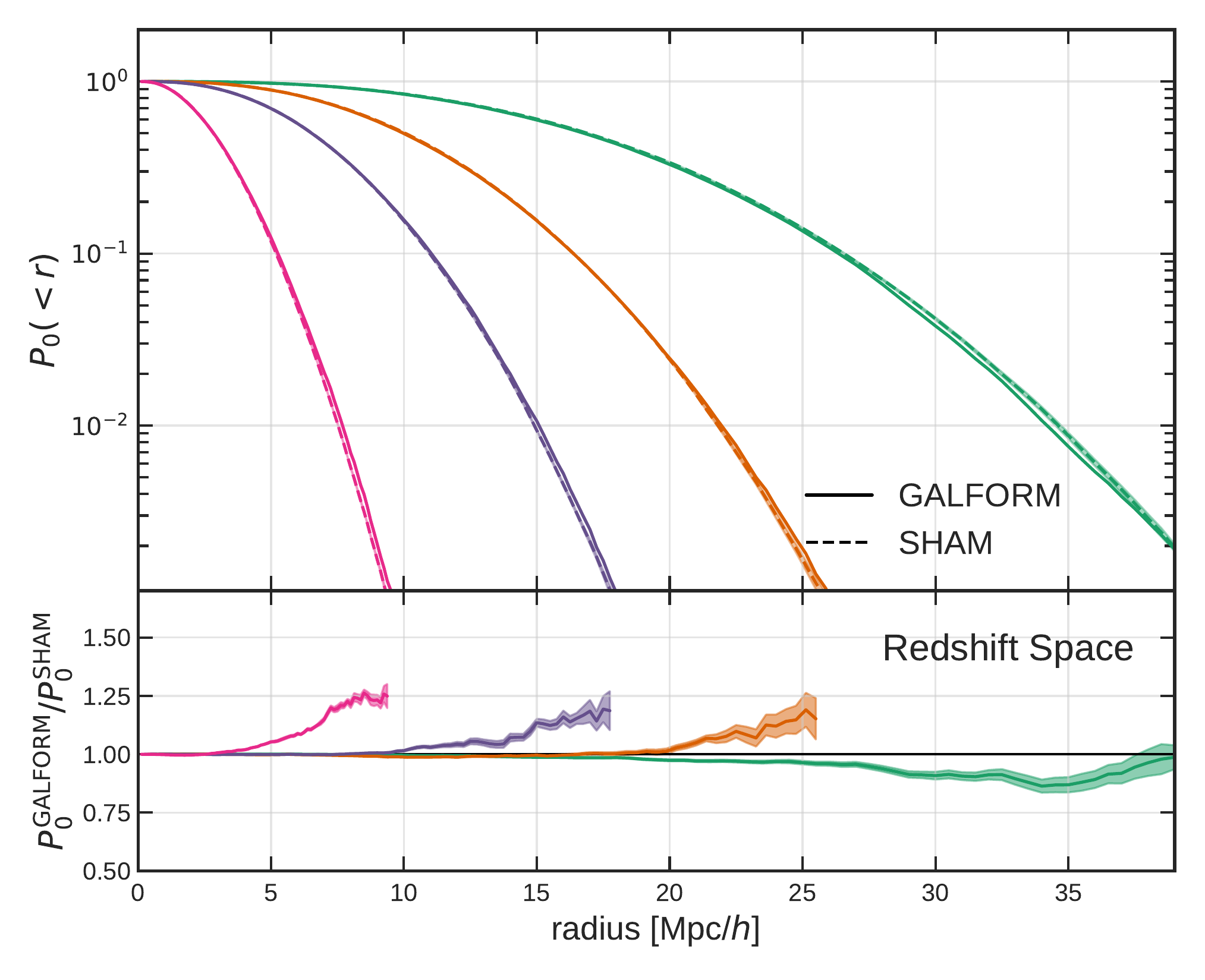}
}
\caption{Void probability function of the 4 samples from \galform\ and SHAM in real space (left panels) and redshift space (right panels). The solid lines show the VPFs of the \galform\ samples, and the dashed lines show the mean VPFs of 10 SHAM realisations (with scatter tuned as discussed in \S\ref{sec:galsham}). The shaded regions show the 1-sigma spread in the VPFs from these 10 SHAM realisations. Each VPF is measured with the same $10^5$ random spheres over the simulation volume.}
\label{fig:vpf}
\end{figure*}

Fig.~\ref{fig:lgals_hod} shows the HODs of galaxy samples with the same number densities as S1, S2, S3, and S8 in Table~\ref{tab:samples} from \lgalaxies, split by centrals and satellites (top panel) and red and blue galaxies (bottom panel). The colour cut used for \lgalaxies is slightly different to the one used for \galform: $(g-r)=-(M_r-20)/33.3 + 0.6$. As in \galform\ (see Fig. \ref{fig:galformHODsplit}), there are features in the \lgalaxies\ HODs that deviate from the standard HOD form. These arise largely from blue central galaxies, although the sharp feature seen in S3 in \galform\ is not present in \lgalaxies. Similarly to the \galform\ HODs, the occupation of blue galaxies is not monotonic in all but the brightest of the \lgalaxies\ samples (S2, S3, and S8). However, the top panel shows that the occupation of central galaxies in \lgalaxies\ does plateau at $\langle N \rangle =1$ as is expected in the standard HOD model, which is not the case for all of the \galform\ samples.

Fig.~\ref{fig:lgals_hod} indicates that while the shapes of the \galform\ HODs may be somewhat extreme, some of the features we noted may be generic of semi-analytic models that include AGN feedback, such as the non-monotonic occupation of blue galaxies.

Because AGN feedback more strongly affects galaxy luminosity than stellar mass, it is interesting to look at the HODs of samples defined by stellar mass as opposed to luminosity in both semi-analytic models. In fact, the standard HOD model was based in part on semi-analytic galaxy samples defined by mass as opposed to luminosity \citep{benson2003}, so one could expect these to have a more standard shape, even though they were developed before the inclusion of AGN feedback in semi-analytic galaxy formation models \citep[e.g.][]{croton2006,bower2006}. 

The top panel of Fig.~\ref{fig:galformlgalsmstar} shows four samples from \galform\ with the same number densities as the samples used previously, but here we rank-order in stellar mass as opposed to luminosity (we refer to these samples with an asterisk to distinguish them from the luminosity-threshold samples). We also show the occupation of central (dashed lines) and satellite galaxies (dot-dashed lines) in these samples. The bottom panel of Fig.~\ref{fig:galformlgalsmstar} shows the same as the top panel, but for stellar-mass threshold samples in \lgalaxies. It is clear from this figure that samples ranked by stellar mass have more standard shapes than those ranked by luminosity. This has been pointed out previously in e.g.\ \citet{contreras2015}. The non-standard features in the HODs of luminosity-threshold samples in \galform\ are not present in these stellar-mass-threshold samples. Also, the occupation of central galaxies in each sample is closer to a step function, as assumed in the standard HOD model.

\subsection{Distinguishing samples with different HODs}
\label{sec:discussion2}
Now that we have established that the HODs of luminosity-threshold samples in semi-analytic models including AGN feedback display non-standard features, it is interesting to ask whether it is possible to distinguish two samples with the same 2-point correlation function, at least on large scales, but different HOD shapes? This would give us a possible way to establish whether the standard HOD model is adequate for describing the HODs of real samples, which are often defined in luminosity-threshold samples, like in \citet{zehavi2011}.

We focus on the void probability function (VPF), which gives the probability of finding zero objects within a sphere of a given radius. We use the publicly available {\tt Corrfunc} code, which contains various OpenMP parallelized clustering measures, to compute the VPF\footnote{The {\tt Corrfunc} code is publicly available at \url{https://github.com/manodeep/Corrfunc}} \citep{manodeep_sinha_2016_55161}. We compare the VPF in \galform\ and SHAM for the four clustering-matched, luminosity-ranked samples. Fig.~\ref{fig:vpf} shows the VPF in real space (left panels) and redshift space (right panels) of the four samples in \galform\ (solid lines) and SHAM (dashed lines). We show the real-space VPFs because we are interested in whether differences exist in the samples, and we show redshift-space VPFs because we want to know if these differences persists when probed with real data. The lower panels in both plots show the ratio of the \galform\ VPF to the SHAM VPF. The shaded regions show the 1-sigma spread in the VPFs from these 10 SHAM realisations. Each VPF is measured with the same $10^5$ random spheres over the simulation volume, reducing Poisson noise in the VPF ratios.

This figure shows that in all four samples, the VPF of \galform\ and SHAM catalogues differ significantly from each other on large scales, even though the 2-point correlation functions on these scales agree. For example, it is clear from this plot that in the faintest sample ($\bar n = 2.3 \times 10^{-2}$ (Mpc/$h$)$^{-3}$, S8), the \galform\ catalogue is more `empty' (higher probability of finding an empty sphere) on scales larger than about $5$ Mpc/$h$. For the brightest sample, $\bar n = 5.0 \times 10^{-5}$ (Mpc/$h$)$^{-3}$ (S1), the SHAM catalogue is more empty on scales larger than about $20$ Mpc/$h$.

Because the 2-point correlation functions of the SHAM samples have been tuned to match that of \galform\ on large scales, the differences in the VPF must arise from either the difference in the HOD shape or from differing levels of assembly bias in the samples. We do not expect the differences in the small-scale 2-point correlation function to significantly impact the VPF on these large scales.

As mentioned previously, we expect both \galform\ and SHAM to have some level of assembly bias inherently. For \galform, this is because the galaxy catalogue is built using the full merger history of the subhaloes. For SHAM, the subhalo quantity that we abundance match on, $\vpeak$, contains information about the merger history of the subhalo. Thus, in order to examine the impact of assembly bias on the VPFs in \galform\ and SHAM, we form shuffled versions of these catalogues, which removes any halo assembly bias present in the original samples. In practice, this involves taking all galaxies (or subhaloes) living in haloes within a given halo mass bin, and randomly shuffling them around to different haloes within that mass bin, keeping satellite galaxies with their centrals. The resulting shuffled samples have exactly the same HOD shape as the original sample, but removes any dependence of the HOD on other halo properties.

\begin{figure}
\centering
\subfloat{
\includegraphics[width=0.48\textwidth]{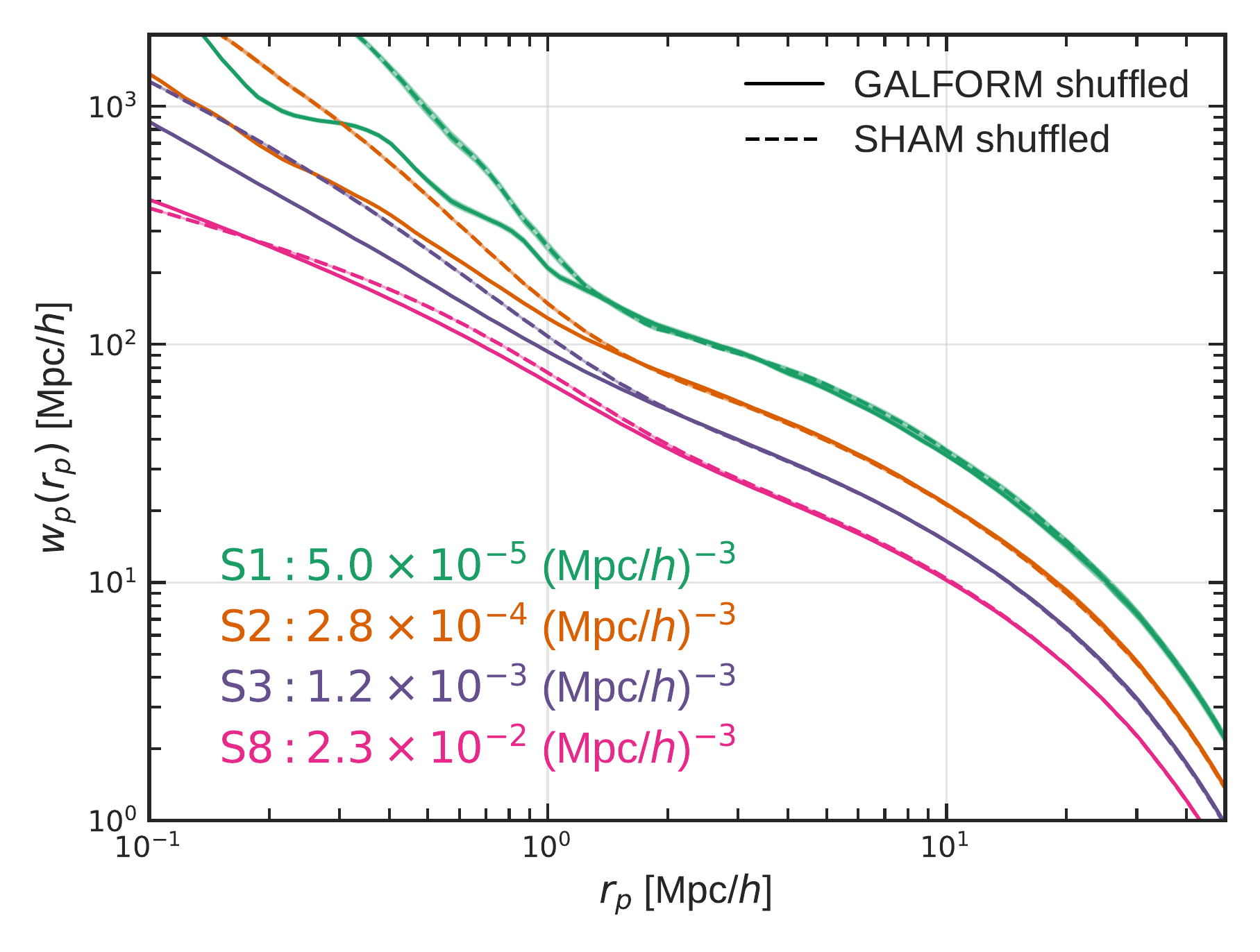}
}
\\
\subfloat {
\includegraphics[width=0.48\textwidth]{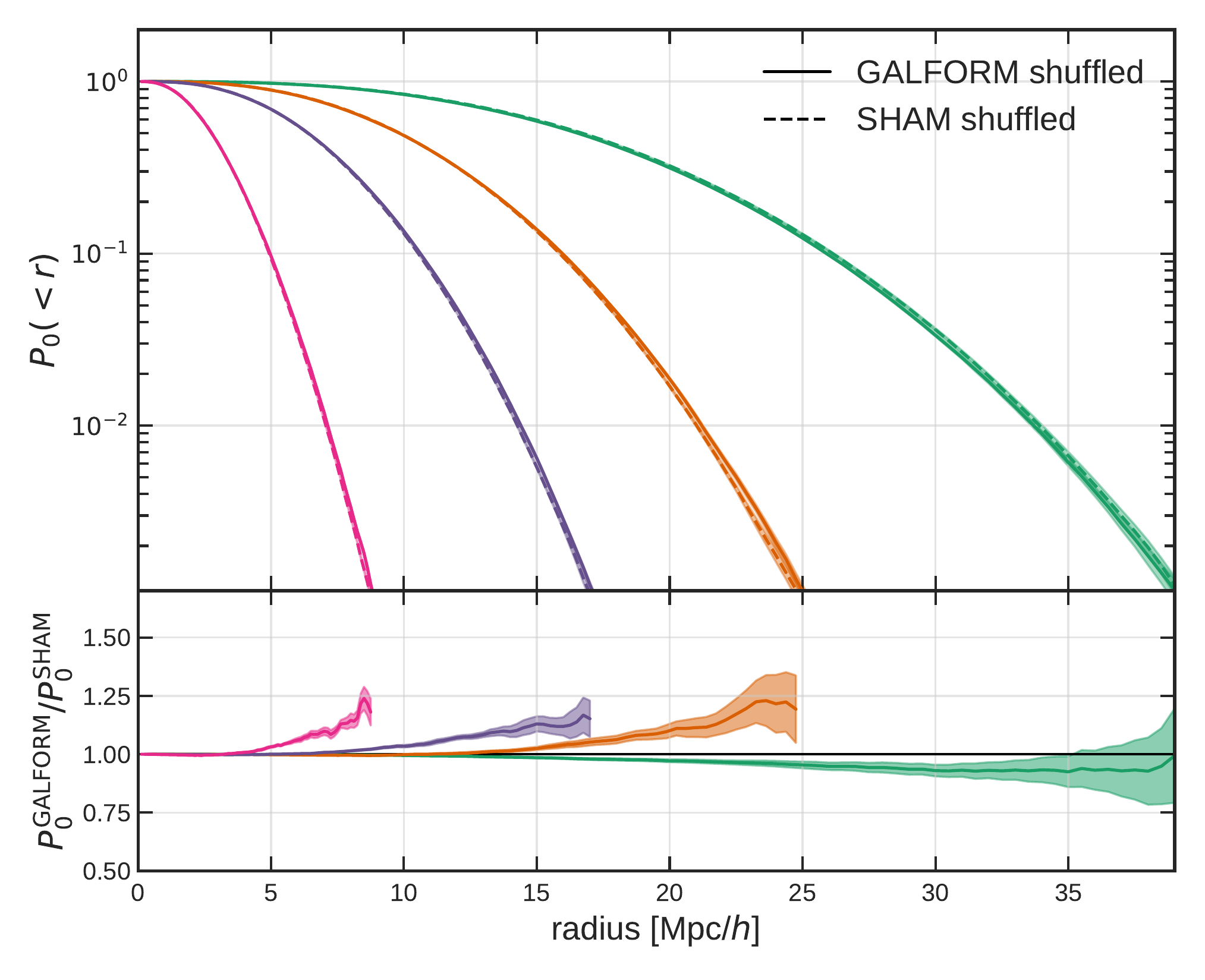}
}
\caption{Top panel: projected correlation functions of shuffled \galform\ and SHAM samples (solid and dashed lines, respectively). Shaded regions show 1-sigma spread from 10 shuffled samples. Bottom panels: void probability functions of shuffled \galform\ and SHAM samples. Bottom-most panel shows the mean ratio between VPFs in all 100 pairs of shuffled \galform\ and SHAM samples. Shaded region shows 1-sigma spread in the ratio. The differences in VPFs here must arise from differences in HOD shapes, rather than assembly bias. For description of shuffling procedure, see \S\ref{sec:discussion2}.}
\label{fig:shuffle}
\end{figure}

Fig.~\ref{fig:shuffle} shows the projected correlation functions (top panel) and VPFs (bottom panels) of the shuffled \galform\ and SHAM samples in real space. For SHAM, we shuffle each of the 10 unique SHAM catalogues once, and the shaded regions around the dashed lines show the spread in those measurements. For \galform, we shuffle the original samples 10 times, and the shaded regions around the solid lines show the spread in the measurements from the shuffled samples. The bias measured from the correlation function of the shuffled samples on large scales agrees statistically with the bias values in Table~\ref{tab:samples} measured from the HODs, as expected. The bottom-most panel of Fig.~\ref{fig:shuffle} shows the ratio between the VPFs. The solid lines show the mean ratio between all 100 combinations of shuffled \galform\ and shuffled SHAM, and the shaded regions show the 1-sigma spread in these. 

It is clear from the bottom panel of Fig.~\ref{fig:shuffle} that even when assembly bias is removed from the \galform\ and SHAM catalogues, differences in the VPFs remain, and show similar trends to the differences we observed in Fig.~\ref{fig:vpf}. From this, we conclude that the differences in the VPFs in Fig.~\ref{fig:vpf} arise from the differences in the HOD shape between \galform\ and SHAM. Because the differences persist in redshift space, in principle the VPF could be used to distinguish two catalogues with similar large-scale 2-point clustering but different HODs. Such a study is left to a future paper.

\section{Conclusions}
\label{sec:conclusion}

We explored various aspects of HOD and SHAM in the context of the semi-analytic galaxy formation models, \galform\ and \lgalaxies. We showed that the standard 5-parameter HOD model (Eq.~\ref{eq:hodmodel}) is not adequate to describe the mean galaxy occupation of haloes in luminosity-threshold samples in these models, whereas it is adequate for describing SHAM samples with the same large-scale clustering. The shortcomings of the standard HOD model may be related to AGN feedback, which complicates the relationship between stellar mass, luminosity, and halo mass. We showed that the HODs of stellar-mass threshold samples in \galform\ and \lgalaxies\ have more standard shapes than luminosity-threshold samples.

We used the VPF to distinguish samples with the same large-scale clustering but different HOD shapes. While the observed differences in VPFs of our samples could arise from assembly bias, we showed that the differences persist even when assembly bias is removed from all samples through shuffling. Thus, we showed we can in principle use the VPF to test whether the HOD model is a good description of a sample.

We also noted that, while much of the HOD fitting done in the literature is presented for a mean occupation of 0.1 and above, features below this level can have a significant impact on the overall number density and bias of the sample, due to the steepness of the halo mass function.

This study should be extended to hydrodynamical simulations once the volumes are large enough for precision measurements on large scales: current clustering studies from state-of-the-art hydrodynamical simulations are restricted to scales below $\sim 5$~\mpch~\citep[e.g.][]{artale2017, ChavesMontero2016}.

\section*{Acknowledgements}
This work was supported by the Science and Technology Facilities Council (ST/L00075X/1). PN acknowledges the support of the Royal Society through the award of a University Research Fellowship, and the European Research Council, through receipt of a Starting Grant (DEGAS-259586). 

This work used the DiRAC Data Centric system at Durham University, operated by the Institute for Computational Cosmology on behalf of the STFC DiRAC HPC Facility (www.dirac.ac.uk). This equipment was funded by BIS National E-infrastructure capital grant ST/K00042X/1, STFC capital grant ST/H008519/1, and STFC DiRAC Operations grant ST/K003267/1 and Durham University. DiRAC is part of the National E-Infrastructure.





\bibliographystyle{mnras}
\bibliography{bibliography}



\appendix
\section{Dust Extinction in GALFORM}
\label{sec:appendix}

We apply a tapering transformation to the r-band extinctions in \galform\ because of the small population of galaxies with unrealistically small sizes, large optical depths, and thus large dust attenuation. The tapering we apply has the functional form:
\begin{equation}
\delta M_r^T(\delta M_r, \tau)=\frac{2\tau}{\pi}\arctan\left(\frac{\pi \delta M_r}{2\tau}\right) \label{eq:tapering}
\end{equation}
where $\delta M_r^T$ is the tapered dust extinction, which is a function of the original dust extinction found in \galform, $\delta M_r$, as well as the desired maximum extinction value, $\tau$. Eq.~\ref{eq:tapering} smoothly tapers the extinctions to the fixed maximum value of $\tau$ while preserving the values for small extinctions.

We find that the tapering does not significantly affect the clustering or HOD of any of our \galform\ samples, as the highly extincted galaxies are always a negligible contribution to the population in any luminosity threshold sample.

In order to taper the extinctions in g-band ($\delta M_g^T$), we fit a power law to the original extinctions in r-band and g-band ($\delta M_r$, $\delta M_g$), and then compute the tapered g-band extinction with the tapered r-band extinction ($\delta M_r^T$) with the parameters of the power-law fit.


\bsp	
\label{lastpage}
\end{document}